\documentclass[10pt,journal]{IEEEtran}
\usepackage{graphicx}
\usepackage{orcidlink}
\usepackage{float}
\usepackage{bm}
\usepackage{subfig}
\usepackage{amsmath,amsfonts}
\usepackage{mathtools}
\usepackage{algorithm}
\usepackage{algorithmic}
\usepackage[normalem]{ulem}
\usepackage[tableposition=top]{caption}
\usepackage{url}
\usepackage{booktabs}

\usepackage{array}
\usepackage{cite}
\usepackage{color}
\usepackage{cleveref}
\usepackage{geometry}
\usepackage{amssymb}
\usepackage{cuted}
\usepackage{lipsum} 
\usepackage{multicol}
\usepackage{amsmath,amssymb,amsthm,bm}
\newtheorem{theorem}{Theorem}
\newtheorem{corollary}{Corollary}
\newgeometry{left=1.4cm,right=1.4cm,bottom=1.7cm,top=1.5cm}

\begin{document}

\title{Space-Time-Frequency Synthetic
Integrated Sensing and Communication Networks}

\author{
Henglin Pu$^{\orcidlink{0009-0008-2390-5722}}$, 
Xuefeng Wang$^{\orcidlink{0000-0002-6536-4206}}$,
Lu Su$^{\orcidlink{0000-0001-7223-543X}}$, \IEEEmembership{Member, IEEE},
Husheng Li$^{\orcidlink{0009-0009-5699-9558}}$, \IEEEmembership{Senior Member, IEEE}

\thanks{Henglin Pu, Lu Su and Husheng Li are with the Elmore Family School of Electrical and Computer Engineering, Purdue University, West Lafayette, Indiana, USA-47907 (email:
 pu36@purdue.edu, lusu@purdue.edu, husheng@purdue.edu).}%
\and
\thanks{Xuefeng Wang and Husheng Li are with the School of Aeronautics and
 Astronautics, Purdue University, West Lafayette, Indiana, USA-47907 (email:
 wang6067@purdue.edu, husheng@purdue.edu).}\and\thanks{This work was supported by the National Science Foundation under grants 2348826, 2343465, 2343469 and 2418106. }
}

\maketitle

\thispagestyle{empty}

\IEEEpeerreviewmaketitle

\begin{abstract}
Integrated sensing and communication (ISAC) promises high spectral and power efficiencies by sharing waveforms, spectrum, and hardware across sensing and data links. Yet commercial cellular networks struggle to deliver fine angular, range, and Doppler resolution due to limited aperture, bandwidth, and coherent observation time. In this paper, we propose a space–time–frequency synthetic ISAC architecture that fuses observations from distributed transmitters and receivers across time intervals and frequency bands. We develop a unified signal model for multistatic and monostatic configurations, derive Cramér–Rao lower bounds (CRLBs) for the estimations of position and velocity. The analysis shows how spatial diversity, multiband operation, and observation scheduling impact the Fisher information. We also compare the estimation performance between a concentrated maximum likelihood estimator (MLE) and a two stage information fusion (TSIF) method that first estimates per-path delay and radial speed and then fuses them by solving a weighted nonlinear least-squares problem via the Gauss-Newton algorithm. Numerical results show that MLE approaches the CRLB in the high signal-to-noise ratio (SNR) regime, while the two stage method remains competitive at moderate to high SNR but degrades at low SNR. A central finding is that fully synthesized network processing is essential, as estimations by individual base stations (BSs) followed by fusion are consistently inferior and unstable at low SNR. This framework offers a practical guidance for upgrading existing communication infrastructure into dense sensing networks.
\end{abstract}


\section{Introduction}

Integrated sensing and communications (ISAC)~\cite{ISAC1,ISAC2,ISAC3} has attracted substantial attention in recent years due to its potential to significantly enhance spectral and power efficiency in wireless systems. It has evolved into one of the major themes for next-generation wireless networks, particularly in the context of 6G systems. ISAC unifies sensing and communication within a single system infrastructure, leveraging shared spectral and hardware resources. This integration simplifies system deployments and enables transformative applications such as environment monitoring~\cite{environment1}, autonomous driving~\cite{autonomous_driving1}, human-computer interaction~\cite{HCI1}, and smart city initiatives~\cite{smartcity1}.

In monostatic sensing, an ISAC transceiver utilizes the reflected electromagnetic (EM) waves to infer information about surrounding reflectors. When a communication receiver performs sensing using the received signal, the configuration is referred to as bi-static sensing. If multiple receivers participate in sensing (cooperating through ISAC’s communication capabilities) the system achieves multi-static sensing. When multiple transmitters jointly illuminate the environment and capture the scattered signals, the setup forms a multiple-input multiple-output (MIMO) radar sensing system. In these configurations, communications and sensing are carried out within the same EM wave emission cycle. Since a single waveform serves both purposes, bandwidth and power consumptions are significantly reduced, enabling the development of cyber–physical systems that demand simultaneous communication and sensing, such as autonomous driving and unmanned aerial vehicle (UAV) network control.

Since ISAC inherently combines radar and communication functions within a shared hardware system, a commercially viable path toward the deployment of ISAC is to leverage existing communication infrastructure. As the global telecom tower market reached 4.93 million units in 2024, with forecasts projecting growth to 5.90 million units by 2033~\cite{IMARC2025TelecomTower}, the massive installed base stations (BSs) offers great potential to augment them with sensing capabilities, thereby enabling a wide range of practical applications. While the commercial benefits of using communication networks for large-scale sensing are clear, a critical challenge lies in the mismatch between radar sensing requirements and the capabilities of commercial communication devices. High-resolution sensing demands large apertures, wide bandwidths, and long observation times, all characteristics that typical communication hardware cannot readily provide. 

\begin{itemize}
    \item Large Aperture for Fine Angular Resolution: Achieving fine angular resolution demands physically large apertures. Yet antennas on commercial base stations are constrained in size, and those on user equipment are even smaller, limiting achievable beamforming gain.
    \item Large Bandwidth for Fine Range Resolution:  High range resolution requires wide signal bandwidths, while current 5G systems typically provide only a few hundred megahertz~\cite{release19}, far below the multi-gigahertz bandwidths needed for sub-meter ranging accuracy. 
    \item Long Time Duration for Fine Doppler Resolution: Similarly, fine Doppler resolution necessitates long coherent observation times. However, communication signals are often bursty and intermittent, making it difficult to sustain the durations required for precise velocity estimation.
\end{itemize}

\begin{figure}[!t]
 \centering
 \includegraphics[width=3.7in]{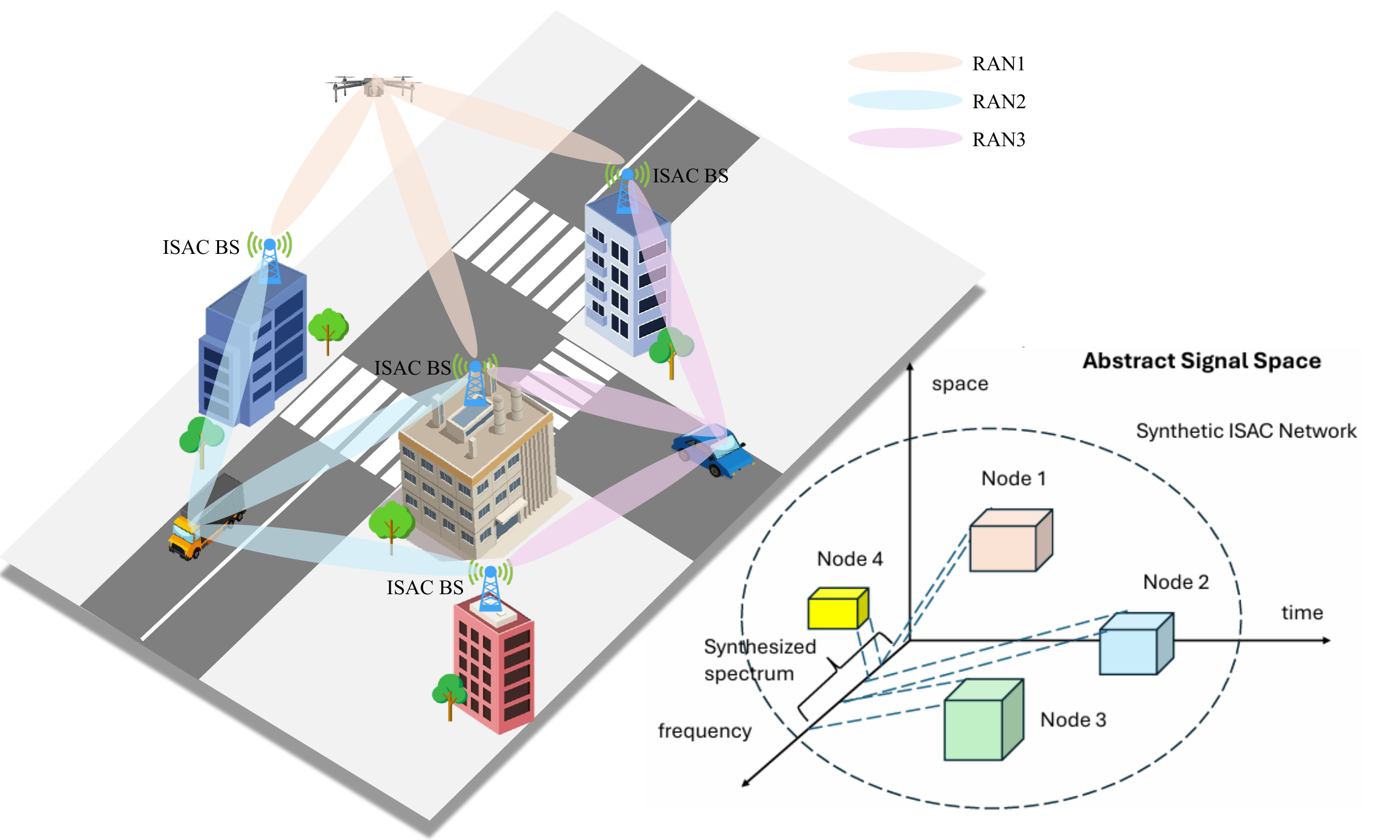}\\
 \caption{Illustration of space-time-frequency synthetic ISAC network}
 \label{fig:illustration}
\end{figure}

Existing ISAC studies typically address only one or two of these dimensions and do not realize a unified space–time–frequency synthetic system. To close this gap, we propose a space–time–frequency synthetic ISAC network, motivated by synthetic aperture radar (SAR), that fuses measurements from multiple communication devices distributed across locations, operating on heterogeneous frequency bands, and over staggered time intervals. An illustration is shown in Fig.~\ref{fig:illustration}. The goal is to assemble a larger virtual sensing aperture by pooling measurements across distributed base stations, across staggered time intervals, and across multiple carrier bands while remaining compatible with commodity communication stacks. In frequency, we synthesize bandwidth by scheduling hops over a set of noncontiguous component carriers. In 5G NR this is naturally supported through synchronization signal blocks (SSB) and carrier aggregation, such that sensing can ride on traffic-friendly resources without hardware changes. In the time domain, we synthesize a longer coherent observation by stitching a slow time pulse train collected over successive frames and slots, with each pulse time referenced and phase referenced so the aggregate coherent processing interval (CPI) behaves like an extended coherent aperture within oscillator and channel coherence limits. In space, we synthesize aperture by deploying multiple base stations and user equipment that illuminate and observe common reflectors from diverse angles. Each transmit receive path contributes delay and Doppler information determined by its geometry, and the network level information is the sum across paths. Coherent synthesis across different BSs, however, is challenging because sensing requires far tighter phase alignment than communication. In large networks with distant base stations a shared phase reference within sensing tolerances is often impractical. We therefore adopt a self-coherent but inter-BS noncoherent model: each BS maintains internal phase coherence across its own pulses and scheduled carriers, whereas measurements from different base stations are combined without assuming a common phase reference. Each BS performs matched filtering and waveform separation, then forwards either complex slow time snapshots for fully synthesized maximum likelihood processing or compact per-path delay and radial speed summaries with covariances for two-stage information fusion (TSIF). This design reuses existing radios and schedulers, scales with the number and placement of BSs, and trades modest temporal and spectral diversity for large synthetic gains in range and Doppler resolution. In summary, our main contributions can be summarized as follows:
\begin{itemize}
    \item We introduce a unified space-time-frequency synthetic ISAC architecture that combines observations from distributed transmitters and receivers across time intervals and frequency bands, and that can be deployed on existing communication infrastructure without hardware changes.
    \item We develop a signal model covering multistatic and monostatic configurations, derive closed-form Cram\'{e}r-Rao Bounds (CRLBs) for joint position and velocity estimations, and quantify how synthesized bandwidth, coherent time, and base station geometry affect the estimation accuracy. The analysis reveals a coupling between delay and velocity induced by frequency hopping, from which we conclude that, under this coupling, increasing synthesized bandwidth should be prioritized over extending  CPI time.
    \item We propose a two-stage information fusion estimator that first estimates per-path delay and radial speed and then fuses them by solving a weighted nonlinear least-squares problem . Benchmarked against a concentrated maximum likelihood estimator (MLE) that operates directly on network-wide raw measurements, the fusion approach is competitive at moderate to high SNR but degrades at low SNR, underscoring the need for fully synthesized processing rather than separate per-BS estimation followed by fusion.
\end{itemize}

The remainder of
this paper is organized as follows. The general
system model is introduced in Section \ref{sec:system_model}. Then, a detailed derivation of CRLBs is elaborated in
Section \ref{sec:CRLB}. The estimation methods including MLE and TSIF are described in Section \ref{sec:estimators}. Then, the numerical results are provided in Section \ref{sec:results}. Finally the
conclusions are drawn in Section \ref{sec:conclusion}.

\section{Related Works}

To mitigate the limitations of communication infrastructure for ISAC, several research directions have been explored. Array-aperture expansion has been pursued via cooperative and distributed ISAC architectures, where multiple spatially separated nodes jointly emulate a larger virtual array. In~\cite{Demirhan2023CFISAC_CellFree}, the authors propose BS-free ISAC beamforming in which distributed APs jointly serve users and sense targets, deriving a multistatic sensing SNR and optimizing joint sensing/communication precoders, a canonical example of spatial synthesis in ISAC. Follow-ups on BS-free ISAC analyze scalable detection/tracking, power control, and multistatic detection under distributed APs, emphasizing coordination requirements and the gains from spatial diversity~\cite{CFISAC_Scalable2025,CFISAC_Survey2025,Silva2023CFisac}. The DISAC vision in~\cite{DISAC2024} argues for distributed, intelligent ISAC with cross-node sensor fusion, which elevates synchronization and calibration to first-class design concerns. In~\cite{Han2025OTASync}, over-the-air time–frequency synchronization is developed for distributed ISAC, exploiting multistatic reciprocity to estimate carrier frequency offset (CFO) without strict line-of-sight (LoS). 

When wide contiguous spectrum is unavailable, frequency-hopping and multiband designs can effectively synthesize wideband observations. Wan \textit{et al.}\cite{Wan2024VTM_MultibandISAC} analyze OFDM-based multiband ISAC, quantifying resolution limits and providing algorithms that stitch non-contiguous bands while managing phase-distortion factors. Tagliaferri \textit{et al.}\cite{Tagliaferri2022DualDomainISAC,Tagliaferri2023TWC} superpose a sensing component onto OFDM in the delay–Doppler (dual) domain to preserve communication compatibility while enhancing range–Doppler resolution. HiSAC~\cite{HiSAC2024} demonstrates a multiband ISAC prototype that achieves super-resolved passive ranging by fusing non-contiguous carriers (e.g., carrier aggregation), yielding 3–20 fold improvements over baselines. A tutorial perspective in~\cite{Koivunen2024SPM} summarizes multicarrier ISAC and systematizes waveform/processing under non-idealities and shared-spectrum constraints, which is useful for designing bandwidth synthesis on commodity stacks. 

Another line of work addresses Doppler resolution under short or bursty transmissions. Islam \textit{et al.}\cite{Islam2022FDmmWaveISAC} propose full-duplex mmWave ISAC with hybrid analog/digital beamforming to sense while transmitting, improving  CPI efficiency at the cost of stringent self-interference suppression. OFDM-ISAC processing under long-range echoes and burstiness has been studied in\cite{Wang2023OFDM_ISI_ICI_ISAC,Mura2023OFDM_ISAC_RO} via coherent-compensation techniques and resource-occupancy–aware waveform design to preserve Doppler and ambiguity performance.

\section{System Model}
\label{sec:system_model}


We consider a distributed ISAC network consisting of $M$ transmitters and $N$ receivers. The network operates in a self-coherent but inter-node noncoherent manner: each ISAC BS maintains internal phase coherence across its own pulses, while no common phase reference is assumed between different BSs. The transmitting and receiving ISAC nodes are located in the two-dimensional plane $(x,y)$. The $M$ transmitters are arbitrarily located at coordinates $\mathbf{t}_k=(x_{tk},y_{tk})$, $k=1,\ldots,M$, and the $N$ receivers are arbitrarily located at $\mathbf{r}_\ell=(x_{r\ell},y_{r\ell})$, $\ell=1,\ldots,N$. The total number of multistatic transmit–receive paths is $L=MN$. In this work, we focus on a
single-target scenario, where the target is located at an unknown position
$\mathbf{x}=\begin{bmatrix}x & y\end{bmatrix}^{\mathsf T}\in\mathbb{R}^2$ and moves with an unknown
constant velocity $\mathbf{v}=\begin{bmatrix}v_x & v_y\end{bmatrix}^{\mathsf T}\in\mathbb{R}^2$.
Extensions to multiple-target configurations are left for future work. We consider both the multistatic and monostatic settings. In the remainder of the paper, we always start with the multistatic case, as it is the more general setting and can be readily specialized to the monostatic case by adjusting the geometry.


\subsection{Target–Path Parameters}
For each transmitter–receiver pair $(k,\ell)$, the multistatic time delay is
\begin{equation}\label{eq:multistatic_delay}
\tau_{k\ell}(\mathbf{x}) 
= \frac{1}{c}\big( \|\mathbf{x} - \mathbf{t}_k\| + \|\mathbf{x} - \mathbf{r}_\ell\| \big),
\end{equation}
where $c$ is the light speed.  The corresponding multistatic Doppler shift is
\begin{equation}\label{eq:multistatic_Doppler}
f_{k\ell}(\mathbf{v}) 
= \frac{f_c}{c}\,\big( \mathbf{u}_{t,k} + \mathbf{u}_{r,\ell} \big)^{T}\mathbf{v},
\end{equation}
where $f_c$ is the carrier frequency, $\mathbf{u}_{t,k}$ is the unit vector pointing from the target 
towards transmitter $k$, and $\mathbf{u}_{r,\ell}$ is the unit vector pointing from the target towards 
receiver $\ell$. Explicitly, $\mathbf{u}_{t,k}$ and $\mathbf{u}_{r,\ell}$ can be written as
\begin{equation}
  \mathbf{u}_{t,k} = 
\begin{bmatrix}\cos\phi_k \\ \sin\phi_k \end{bmatrix}, \qquad
\mathbf{u}_{r,\ell} = 
\begin{bmatrix}\cos\varphi_\ell \\ \sin\varphi_\ell \end{bmatrix}, 
\end{equation}
where $\phi_k$ and $\varphi_\ell$ denote the azimuth angles from the target to the $k$th transmitter 
and $\ell$th receiver, respectively. For convenience we define the path geometry vector
\begin{equation}\label{eq:geometry}
\mathbf{g}_{k\ell} \triangleq \mathbf{u}_{t,k} + \mathbf{u}_{r,\ell},
\end{equation}
which will be used for later derivation of CRLB.

\subsection{Transmit Waveforms}

In line with practical ISAC deployments, we assume that each transmitter $k$ employs a multicarrier communication waveform, specifically OFDM. Each transmitter $k$ emits a baseband waveform $s_k(t)$, assumed medium bandwidth with effective 
(root-mean-square) bandwidth $\beta_k$. Let $S_k(f)$ denote the Fourier transform of $s_k(t)$. The effective bandwidth $\beta_k$ is defined as~\cite{radar_book}
\begin{equation}
    \beta_k^2 = \frac{\int_{W_k} f^2 |S_k(f)|^2  df}{\int_{W_k} |S_k(f)|^2  df},
\end{equation}
where $W_k$ is the occupied band. The waveforms are assumed to be mutually orthogonal,
\begin{equation}
   \int s_k(t)\, s_m^*(t)\,dt = 0, \quad k \neq m, 
\end{equation}
so that after matched filtering at the receivers the returns corresponding to different transmitters 
can be separated without cross terms. This assumption significantly simplifies the Fisher iInformation 
matrix (FIM) structure, as shown later in the derivations.

We use “fast time” for the within-pulse time used for range sampling, and “slow time” for the pulse index across a CPI. Within a CPI of $P_k$ slow-time pulses ($p=1,\dots,P_k$ with nominal
times $\{t_{k,p}\}$ and pulse repetition interval (PRI) $T_
{r,k}$), transmitter $k$ uses a frequency-hopped carrier
sequence $f^{k}_{c,p}$ around its nominal band. Let $s^k_p(t)$
correspond to an OFDM sensing block, typically constructed from predesigned pilot symbols (and, if
desired, known communication symbols) assigned to transmitter $k$ in the $p$-th hop. The resulting baseband transmit signal over one CPI can be written as
\begin{equation}
    s_k(t)
     = 
    \sum_{p=1}^{P_k} s^k_p\!\big(t-pT_r\big)
    \,e^{j 2\pi f^{k}_{c,p}\, t}.
    \label{eq:fh-synthesis}
\end{equation}
Each hop is narrowband, while the union of $\{f^{k}_{c,p}\}_{p=1}^{P_k}$ synthesizes a much larger
effective bandwidth across frequency. This synthesis occurs over multiple slow-time pulses, thereby
also enlarging the effective coherent observation interval in time and enhancing the velocity (Doppler)
estimation capability naturally.
Note that with the frequency-hopping scheme, in each time slot the BSs occupy disjoint OFDM subcarrier sets within nonoverlapping bands. Consequently, the transmit waveforms have disjoint spectral support and are orthogonal under the fast–time matched filter.

\subsection{Received Signal Model}
For each path $(k,\ell)$ we collect $P_{k\ell}$ pulses in slow time, transmitted at instants 
$t_{k\ell,p}$, $p = 0,\ldots,P_{k\ell}-1$ ($t_{k\ell,p}=p T_{r,k}$ with PRI $T_{r,k}$). After matched filtering to the orthogonal waveform $s^k_p(t)$, the complex baseband received signal 
for the $(k,\ell)$th path at the $p$th pulse is given by
\begin{equation}\label{eq:received_signal}
   y_{k\ell}[p] = \alpha_{k\ell}\, s^k_p(\tau_{k\ell})
e^{-j2\pi f^{k\ell}_{c,p}\tau_{k\ell}}e^{j 2\pi f^{k\ell}_{D,p} t_{k\ell,p}} + w_{k\ell}[p], 
\end{equation}
where the Doppler shift is given by $f^{k\ell}_{D,p} = \frac{f^{k\ell}_{c,p}}{c}\mathbf{u}_{k\ell}^T\mathbf{v}$, $\alpha_{k\ell}$ is an unknown complex path gain, $f^{k\ell}_{c,p}$ denotes the central frequency at $t_{k\ell,p}$ and $w_{k\ell}[p]\sim \mathcal{CN}(0,\sigma_w^2)$ is circular complex Gaussian noise, independent across sensors and pulses. Within each base station we assume a coherent processing interval, so that all slow time phase evolution is captured explicitly by the delay–frequency term $e^{-j2\pi f^{k\ell}_{c,p}\tau_{k\ell}}$ and the Doppler term $e^{j 2\pi f^{k\ell}_{D,p} t_{k\ell,p}}$. The remaining BS-dependent phase offsets (local oscillator (LO) and radio-frequency (RF) chain phase and target reflection phase) do not vary with $p$ and are absorbed into the constant complex gain $\alpha_{k\ell}=|\alpha_{k\ell}|e^{j\phi_{k\ell}}$.


\section{CRLBs for Sensing}\label{sec:CRLB}

The CRLB provides a lower bound for the mean squared error (MSE) of any unbiased estimator for an unknown parameter(s). With observation vector $\mathbf{r}$ and likelihood $p(\mathbf{r}\mid \boldsymbol{\theta})$, the Fisher information matrix (FIM) is given by
\begin{equation}
\mathbf{J}(\boldsymbol{\theta})  =  
E_{\mathbf{r}\mid \boldsymbol{\theta}}\!\left\{
\left[ \frac{\partial}{\partial\boldsymbol{\theta}}\log p(\mathbf{r}\mid\boldsymbol{\theta})\right]
\left[\frac{\partial}{\partial\boldsymbol{\theta}}\log p(\mathbf{r}\mid\boldsymbol{\theta})\right]^{\!T}
\right\},
\end{equation}
where $E_{\mathbf{r}\mid \boldsymbol{\theta}}\{\cdot\}$ denotes conditional expectation. 
The CRLB matrix is then defined as
\begin{equation}
\mathbf{C}_{\text{CRLB}}  =  \mathbf{J}^{-1}(\boldsymbol{\theta}).
\end{equation}


Sometimes, it is easier to compute the FIM with respect to another vector $\boldsymbol{\eta}$, and apply the chain rule to derive the original $\mathbf{J}(\boldsymbol{\theta})$. In our case, since the received signals (\ref{eq:received_signal}) are functions of the time delays $\tau_{k\ell}$, the Doppler shift $f_{k\ell}$ and the complex amplitudes, by the chain rule, $\mathbf{J}(\boldsymbol{\theta})$ can be expressed in the alternative form~\cite{Kay_book}
\begin{equation}\label{eq:chain_rule}
    \mathbf{J}(\boldsymbol{\theta}) = \mathbf{P} \mathbf{J}(\boldsymbol{\eta}) \mathbf{P}^{T},
\end{equation}
where $\boldsymbol{\eta}$ is a vector of unknown parameters, and it incorporates the time delays and Doppler shifts. Matrix $\mathbf{J}(\boldsymbol{\eta})$ is the FIM with respect to $\boldsymbol{\eta}$, and matrix $\mathbf{P}$ is the Jacobian
\begin{equation}
   \mathbf{P} = \frac{\partial \boldsymbol{\eta}}{\partial \boldsymbol{\theta}}. 
\end{equation}
In the subsequent discussion, we develop the CRLB for the case of multistatic and monostatic settings, separately.

\subsection{CRLBs under Multistatic Setting}\label{sec:CRLB_multistatic}


The vector of unknown parameters is defined as
\begin{equation}
\boldsymbol{\theta} = 
\begin{bmatrix}
\mathbf{x} & \mathbf{v} & \Re\{\boldsymbol{\alpha}\} & \Im\{\boldsymbol{\alpha}\}
\end{bmatrix}^{T},
\end{equation}
where $\boldsymbol{\alpha}=[\alpha_{11},\ldots,\alpha_{MN}]^{T}$ stacks the complex amplitudes. $\Re\{\dot\}$ and $\Im\{\dot\}$ denote the real and imaginary parts of a complex-valued vector/matrix. Note that we represent the complex gains by their real and imaginary parts so the parameter vector is real, which makes the FIM real and allows us eliminate the nuisance gains via a Schur complement. We first parameterize the unknowns by delays rather than positions and then use the chain rule given in (\ref{eq:chain_rule}) to map the Fisher information back to position estimation, which simplifies the derivations. The intermediate parameters is now given by
\begin{equation}
\boldsymbol{\eta} = 
\begin{bmatrix}
\boldsymbol{\tau} & \mathbf{v} & \Re\{\boldsymbol{\alpha}\} & \Im\{\boldsymbol{\alpha}\}
\end{bmatrix}^{T},
\end{equation}
where $\boldsymbol{\tau} = [\tau_{11},\ldots,\tau_{MN}]^{T}$ stacks the $L$ delays.

\subsubsection{Per-Path Fisher Information}

We first derive the per-path FIM of the observable quantities (delay and velocity) 
with respect to the intermediate parameters. 

For the $k$th transmitter and $\ell$th receiver path, the multistatic delay is given in (\ref{eq:multistatic_delay}). The gradient of $\tau_{k\ell}(\mathbf{x})$ with respect to $\mathbf{x}$ can be written as
\begin{equation}
\nabla_{\mathbf{x}} \tau_{k\ell}(\mathbf{x})
= \frac{1}{c}\left(\mathbf{u}_{t,k} + \mathbf{u}_{r,\ell}\right).
\end{equation}
The Doppler shift associated with the same path is given in (\ref{eq:multistatic_Doppler}). The gradient of $f_{k\ell}(\mathbf{v})$ with respect to $\mathbf{v}$ is
\begin{equation}
\nabla_{\mathbf{v}} f_{k\ell}(\mathbf{v})
= \frac{f_c}{c}\left(\mathbf{u}_{t,k} + \mathbf{u}_{r,\ell}\right).
\end{equation}
Both derivatives share the same geometric vector $\mathbf{g}_{k\ell}$ in (\ref{eq:geometry}). Specifically, we have
\begin{equation}\label{eq:partial_derivatives_location_velocity}
\nabla_{\mathbf{x}} \tau_{k\ell}(\mathbf{x})
= \frac{1}{c}\,\mathbf{g}_{k\ell}, 
\qquad
\nabla_{\mathbf{v}} f_{k\ell}(\mathbf{v})
= \frac{f_c}{c}\,\mathbf{g}_{k\ell}.
\end{equation}

This structural symmetry plays a central role in simplifying the FIM as the localization and velocity information matrices inherit the same geometric weighting 
$\mathbf{g}_{k\ell}\mathbf{g}_{k\ell}^T$, differing only in their signal-dependent 
scaling factors. 
As a result, the final CRLBs for position and velocity naturally share the same geometry matrix $\mathbf{g}_{k\ell}$ in (\ref{eq:geometry}), 
decoupled from the per-path signal strength and processing gain.

We now derive the Fisher information contributed by a single transmit-receive path. The
network-level Fisher information then follows by summation over paths owing to waveform
orthogonality. Consider an arbitrary path ($k$ and $\ell$ index suppressed for brevity) with complex
baseband observation model:
\begin{equation}
\mathbf{y} = \boldsymbol{\mu}(\boldsymbol{\eta}) + \mathbf{w}, 
\qquad 
\mathbf{w}\sim\mathcal{CN}\!\left(\mathbf{0},\,\sigma_w^2\mathbf{I}\right),
\end{equation}
where, with slight abuse of notation, we denote the per–path parameter vector by $\boldsymbol{\eta}$ defined as
\begin{equation}\label{eq:per_path_parameter_vector}
    \boldsymbol{\eta} \triangleq  
\begin{bmatrix}
\tau & \mathbf{v} & \Re\{\alpha\} & \Im\{\alpha\}
\end{bmatrix}^{\!T}.
\end{equation}
Let $\mathbf{s}(\tau)\in\mathbb{C}^{N_t}$ denote the matched filter
fast-time vector (samples around the correlation peak). Stacking slow and fast time, the mean is separable:
\begin{equation}\label{eq:psi}
\begin{aligned}
  &\boldsymbol{\mu}(\boldsymbol{\eta}) =
\alpha \left[ (\mathbf{d}(\mathbf{v}) \odot \mathbf{a}(\tau)) \otimes \mathbf{s}(\tau) \right],
\\
&\mathbf{d}(\mathbf{v}) = [e^{j2\pi f_{D,0}t_0}, \ldots, e^{j2\pi f_{D,P-1}t_{P-1}}]^T, \\
& \mathbf{a}(\tau) = [e^{-j2\pi f_{c,0}\tau}, \ldots, e^{-j2\pi f_{c,P-1}\tau}]^T.
\end{aligned}
\end{equation}
We adopt the standard complex-Gaussian FIM for real parameters:
\begin{equation}
\label{eq:FIM_complex}
\big[J(\boldsymbol{\eta})\big]_{ij}
= 
\frac{2}{\sigma_w^2}\,
\Re\!\left\{
\left(\frac{\partial\boldsymbol{\mu}}{\partial\eta_i}\right)^{\!H}
\left(\frac{\partial\boldsymbol{\mu}}{\partial\eta_j}\right)
\right\}.
\end{equation}

Define the fast-time energy and its delay-derivative norm as
\begin{equation}
E_s \triangleq \|\mathbf{s}(\tau)\|_2^2, 
\qquad 
\|\partial_\tau\mathbf{s}(\tau)\|_2^2 = 4\pi^2\beta^2\,E_s,
\end{equation}
where $\beta$ is the (rms) effective bandwidth of the transmitted waveform.
For slow time, define the sums
\begin{equation}
S_0 = \sum_{p=0}^{P-1} 1 = P,\quad 
S_1 = \sum_{p=0}^{P-1} t_p,\quad
S_2 = \sum_{p=0}^{P-1} t_p^2.
\end{equation}
Then the centered variance can be written as
\begin{equation}
\operatorname{Var}_t  \triangleq  S_2 - \frac{S_1^2}{S_0}
 =  \sum_{p=0}^{P-1} (t_p-\bar t)^2,\qquad \bar t=\frac{S_1}{S_0}.
\end{equation}
We also define the carrier moments
\begin{equation}
F_1  =  \sum_{p=0}^{P-1} f_{c,p},\quad 
F_2  =  \sum_{p=0}^{P-1} f_{c,p}^2,\quad
\operatorname{Var}_f  =  \frac{F_2}{P} - \left(\frac{F_1}{P}\right)^2,
\end{equation}
and mixed sequence $z_p = t_pf_{c,p}$ with variance $\operatorname{Var}_z = \frac{1}{P}\sum_{p=0}^{P-1} z_p^2 - (\frac{1}{P}\sum_{p=0}^{P-1} z_p)^2$. Given (\ref{eq:psi}) and the product rule, we have
\begin{align}
&\frac{\partial\boldsymbol{\mu}}{\partial\,\Re\{\alpha\}} = (\mathbf{d}(\mathbf{v}) \odot \mathbf{a}(\tau)) \otimes \mathbf{s}(\tau),\label{eq:FIM_1}
\\
&\frac{\partial\boldsymbol{\mu}}{\partial\,\Im\{\alpha\}} = j(\mathbf{d}(\mathbf{v}) \odot \mathbf{a}(\tau)) \otimes \mathbf{s}(\tau),\label{eq:FIM_2}\\
&\frac{\partial\boldsymbol{\mu}}{\partial\tau} = 
\alpha\,\Big[(\mathbf{d}(\mathbf{v})\odot\mathbf{a}(\tau))\otimes\partial_{\tau}\mathbf{s}(\tau)
 + (\mathbf{d}(\mathbf{V})\odot\partial_{\tau}\mathbf{a}(\tau))\otimes \mathbf{s}(\tau)\Big], \label{eq:FIM_3}
\\&
\frac{\partial\boldsymbol{\mu}}{\partial \mathbf{v}} = 
\alpha\,\Big[(\partial_{\mathbf{v}}\mathbf{d}(\mathbf{v})\odot\mathbf{a}(\tau))\otimes \mathbf{s}(\tau)\Big].  \label{eq:FIM_4}
\end{align}

Applying \eqref{eq:FIM_complex} and using Kronecker identities, we can obtain

\emph{(i) Amplitude--amplitude block (real/imaginary basis):}
\begin{equation}
\label{eq:Jaa}
\mathbf{J}_{\alpha\alpha}
 = 
\frac{2}{\sigma_w^2}\,S_0 E_s\,\mathbf{I}_2
 = 
\frac{2 P E_s}{\sigma_w^2}\,\mathbf{I}_2,
\end{equation}
where $\mathbf{I}_{2}$ denotes the $2\times2$ identity matrix.

\emph{(ii) Delay--delay element:}
\begin{equation}
\label{eq:Jtautau}
J_{\tau\tau}
 = 
\frac{8\pi^2}{\sigma_w^2}\,|\alpha|^2\,E_s\,(\beta^2\,P+F_2).
\end{equation}

\emph{(iii) Velocity--velocity element:}
\begin{equation}
\label{eq:Jff}
\mathbf{J}_{\mathrm{vv}} = \frac{8\pi^{2}}{\sigma_{w}^{2}} |\alpha|^{2} E_{s} \sum_{p} \left( \frac{f_{c,p}t_{p}}{c} \right)^{2}  \mathbf{g} \mathbf{g}^{T}.
\end{equation}

\emph{(iv) Delay--amplitude, velocity--amplitude cross terms:}

\begin{equation}
\label{eq:Jtau_alpha_zero}
\begin{aligned}
   &\mathbf{J}_{\tau \alpha} = \frac{4 \pi E_s}{\sigma_w^2} F_1 [-\Im\{\alpha^*\}, -\Re\{\alpha^*\}],
\\
&\mathbf{J}_{\mathbf{v}\alpha} = \frac{4 \pi E_s}{\sigma_w^2} S_1 \begin{bmatrix} \Im\{\alpha\} \\ -\Re\{\alpha\} \end{bmatrix} \mathbf{g}^T.
\end{aligned}
\end{equation}

The delay-amplitude and velocity-amplitude cross blocks of the FIM can be eliminated without loss of information by an invertible reparameterization that recenters the
carrier and the slow-time axis. 

\paragraph{Carrier Centering (remove mean carrier from the delay phase)}
Define the mean carrier
\begin{equation}
\bar f_c  \triangleq  \frac{1}{P}\sum_{p=0}^{P-1} f_{c,p},
\end{equation}
and reparameterize
\begin{equation}
\tilde{\alpha} \triangleq \alpha\,e^{-j2\pi \bar f_c\,\tau},
\qquad
\tilde a_p(\tau) \triangleq e^{-j2\pi\,(f_{c,p}-\bar f_c)\,\tau}.
\label{eq:carrier_centering}
\end{equation}
This mapping is bijective and leaves the mean vector unchanged:
$\alpha\,a_p(\tau)=\tilde\alpha\,\tilde a_p(\tau)$ for all $p$.
In the new coordinates,
\(
\sum_p (f_{c,p}-\bar f_c)=0
\)
so that $F_1=0$ and therefore the delay–amplitude cross block vanishes:
\begin{equation}
\,\mathbf J_{\tau \alpha}= \mathbf 0\,.
\end{equation}

\paragraph{Slow-time Centering (shift the time origin)}
Define the slow-time centroid
\begin{equation}
\bar t  \triangleq  \frac{1}{P}\sum_{p=0}^{P-1} t_p,
\end{equation}
and shift the sampling instants
\begin{equation}
\tilde t_p  \triangleq  t_p-\bar t,
\qquad
\tilde d_p(\mathbf v) =  e^{j2\pi f_{D,p}\tilde t_p},
\qquad
\tilde\alpha \leftarrow \tilde\alpha\,e^{j2\pi f_{D,\mathrm{ref}}\bar t},
\label{eq:time_centering}
\end{equation}
where $f_{D,\mathrm{ref}}$ is any reference Doppler (the common factor is absorbed into
$\tilde\alpha$). This reparameterization is also bijective and preserves the likelihood.
It enforces $\sum_p \tilde t_p=0$, i.e., $S_1=0$, which removes the velocity–amplitude
cross block:
\begin{equation}
\,\mathbf J_{\mathbf v \alpha}= \mathbf 0\,.
\end{equation}

\emph{(v) Delay--velocity cross terms:}

\begin{equation}\label{eq:delay_velocity_cross_term}
    \mathbf{J}_{\tau v} = - \frac{8 \pi^2}{\sigma_w^2} |\alpha|^2 E_s \frac{1}{c} \left( \sum_p t_p f_{c,p}^2 \right) \mathbf{g}^T.
\end{equation}
The detailed derivatives of the FIM (\ref{eq:Jaa})–(\ref{eq:Jtau_alpha_zero}) and (\ref{eq:delay_velocity_cross_term}) are given in Appendix \ref{appendix:FIM}.

\subsubsection{Intermediate FIM After Eliminating \texorpdfstring{$\alpha$}{alpha}}
\label{subsec:intermediate_fim_after_alpha}

We now eliminate the unknown complex amplitudes via a Schur complement and assemble the
intermediate FIM across all paths. As in the prior subsection,
we consider an arbitrary path first (indices suppressed), then extend to the full network
by summation over paths. The orthogonality of the transmitted waveforms implies that the
per-path FIMs add without cross-terms, which is proved in Appendix \ref{appendix:orthogonality_proof}.

Given the per-path parameter vector in (\ref{eq:per_path_parameter_vector}), the per-path FIM in block form is given by
\begin{equation}
    \mathbf{J}(\boldsymbol{\eta})
=
\begin{bmatrix}
J_{\tau\tau} & \mathbf{J}_{\tau \mathbf{v}} & \mathbf{J}_{\tau\alpha} \\[2pt]
\mathbf{J}_{\mathbf{v}\tau} & \mathbf{J}_{\mathbf{vv}} & \mathbf{J}_{\mathbf{v}\alpha} \\[2pt]
\mathbf{J}_{\alpha\tau} & \mathbf{J}_{\alpha \mathbf{v}} & \mathbf{J}_{\alpha\alpha}
\end{bmatrix}.
\end{equation}
The amplitude-eliminated information in the \((\tau,\mathbf{v})\) subspace is given by the
Schur complement
\begin{equation}
\label{eq:schur_elimination}
\begin{aligned}
    J_{(\tau,\mathbf{v})\mid\alpha}
 &= 
\begin{bmatrix}
J_{\tau\tau} & \mathbf{J}_{\tau \mathbf{v}}\\[1pt]
\mathbf{J}_{\mathbf{v}\tau} & \mathbf{J}_{\mathbf{vv}}
\end{bmatrix}
-
\begin{bmatrix}
\mathbf{J}_{\tau\alpha}\\[1pt]
\mathbf{J}_{\mathbf{v}\alpha}
\end{bmatrix}
\mathbf{J}_{\alpha\alpha}^{-1}
\begin{bmatrix}
\mathbf{J}_{\alpha\tau} & \mathbf{J}_{\alpha \mathbf{v}}
\end{bmatrix}
\end{aligned}
\end{equation}

Substituting (\ref{eq:Jaa})-(\ref{eq:delay_velocity_cross_term}) into
\eqref{eq:schur_elimination} yields:
\begin{equation}
    \begin{aligned}
        J_{\tau\mid\alpha}
&=
\frac{8\pi^{2}}{\sigma_{w}^{2}} |\tilde{\alpha}|^{2} E_{s} P \left( \beta^{2} + \mathrm{Var}_{f} \right),
\\
\mathbf{J}_{\mathbf{v}\mid\alpha}
&
= \frac{8\pi^{2}}{\sigma_{w}^{2}} |\tilde{\alpha}|^{2} E_{s} \frac{P}{c^{2}} (\mathrm{Var}_{z}+\mathbb{E}[z]^2 ) \mathbf{g} \mathbf{g}^{T}.
    \end{aligned}
\end{equation}
Even after centering \( F_1 = 0 \), \( S_1 = 0 \), a FH-induced \(\tau\)-velocity coupling generally remains:

\begin{equation}
    \begin{aligned}
       \mathbf{J}_{\tau v | \alpha} = -\frac{8\pi^{2}}{\sigma_{w}^{2}} |\hat{\alpha}|^{2} E_{s} \frac{P}{c} \operatorname{Cov}(t, f_{c}^{2}) \mathbf{g}^{T},\\
       \operatorname{Cov}(t, f_{c}^{2}) = \frac{1}{P} \sum_{p} t_{p} f_{c,p}^{2} \quad \text{when } \bar{t} = 0.
    \end{aligned}
\end{equation}

For later compactness, we define the per-path signal weights
\begin{equation}\label{eq:per_path_weights}
    \begin{aligned}
        w^{(\tau)}
&\triangleq \frac{8\pi^2}{\sigma_w^2}\,|\alpha|^2\,PE_s\,(\beta^2+\operatorname{Var}_f), \\
w^{(\mathbf{v})}
&\triangleq \frac{8\pi^2}{\sigma_w^2}\,|\alpha|^2\,E_s\,\frac{P}{c^2}\operatorname{Var}_z, \\
w^{(\times)} &\triangleq -\frac{8\pi^2}{\sigma_w^2}\,|\alpha|^2 E_{s}\,\frac{P}{c}\,\mathrm{Cov}(t,f_c^2).
    \end{aligned}
\end{equation}
\subsubsection{Assembly FIM across paths}
Let \(\mathcal{L}=\{(k,\ell): k=1,\ldots,M, \ell=1,\ldots,N\}\) index the \(L=MN\) paths, and
stack the intermediate parameters as
\(
\boldsymbol{\eta}
=
[\,\boldsymbol{\tau},\,\mathbf{v}\,]^{T}
\),
with \(\boldsymbol{\tau}=[\tau_{k\ell}]_{(k,\ell)\in\mathcal{L}}\) and
\(\mathbf{v}=[v_{k\ell}]_{(k,\ell)\in\mathcal{L}}\).
Waveform orthogonality implies that the intermediate FIM after eliminating the amplitudes is
block-diagonal across paths and between delay and Doppler:
\begin{equation}
\label{eq:J_eta_after_alpha}
\begin{aligned}
    \mathbf{J}_{\boldsymbol{\eta}\mid\boldsymbol{\alpha}}
&=
\begin{bmatrix}
\mathbf{J}_{\boldsymbol{\tau}\boldsymbol{\tau}\mid\boldsymbol{\alpha}} & \mathbf{J}_{\boldsymbol{\tau}\mathbf{v}\mid\boldsymbol{\alpha}}\\[3pt]
\mathbf{J}_{\mathbf{v}\boldsymbol{\tau}\mid\boldsymbol{\alpha}} & \mathbf{J}_{\mathbf{v}\mathbf{v}\mid\boldsymbol{\alpha}}
\end{bmatrix},
\\
\mathbf J_{\boldsymbol{\tau}\boldsymbol{\tau}|\alpha} &= \mathrm{diag}\!\Big(\,w^{(\tau)}_{k\ell}\,\Big)_{(k,\ell)\in L} \in \mathbb{R}^{L\times L},\\
\mathbf J_{\mathbf v\mathbf v|\alpha} &= \sum_{(k,\ell)\in L} w^{(v)}_{k\ell}\ \mathbf g_{k\ell}\mathbf g_{k\ell}^{\!T} \in \mathbb{R}^{2\times 2},\\
\mathbf J_{\boldsymbol{\tau}\mathbf v|\alpha} &= 
\begin{bmatrix}
w^{(\times)}_{k\ell}\ \mathbf g_{k\ell}^{\!T}
\end{bmatrix}_{(k,\ell)\in L} \in \mathbb{R}^{L\times 2}.
\end{aligned}
\end{equation}
Here \(w^{(\tau)}_{k\ell}\), \(w^{(f)}_{k\ell}\) and \(w^{(\times)}_{k\ell}\) are obtained from
\eqref{eq:per_path_weights} using the corresponding values of the \((k,\ell)\)th path. Thus, the intermediate FIM after amplitude elimination is
a diagonal matrix in the \((\boldsymbol{\tau},\mathbf{v})\) coordinates, with entries
that factor cleanly into a signal-dependent weight.

\subsubsection{Chain Rule}
\label{subsec:map_intermediate_to_physical}

The intermediate FIM $\mathbf{J}_{\boldsymbol{\eta}\mid\boldsymbol{\alpha}}$ derived in 
\eqref{eq:J_eta_after_alpha} quantifies information in terms of the 
path-specific delay and velocity parameters. Our ultimate goal, however, is to obtain CRLBs for 
the physical parameters of interest, namely the target position 
$\mathbf{x}=[x  y]^T$ and velocity $\mathbf{v}=[v_x  v_y]^T$. The mapping $\boldsymbol{\eta}\mapsto(\mathbf{x},\mathbf{v})$ is 
deterministic, and the FIM transforms under reparameterization via the chain rule:
\begin{equation}
\label{eq:chain_rule_FIM}
\mathbf{J}_{\mathbf{x},\mathbf{v}}
= \mathbf{G}^T\,\mathbf{J}_{\boldsymbol{\eta}\mid\boldsymbol{\alpha}}\,\mathbf{G},
\end{equation}
where 
\begin{equation}
    \mathbf{G}  \triangleq  \frac{\partial \boldsymbol{\eta}}
{\partial [\,\mathbf{x}^T \mathbf{v}^T\,]}
=
\begin{bmatrix}
\frac{\partial \boldsymbol{\tau}}{\partial \mathbf{x}} & \mathbf{0}\\[3pt]
\mathbf{0} & 1
\end{bmatrix}.
\end{equation}

The partial derivatives of delay of each path with respect to location are given in (\ref{eq:partial_derivatives_location_velocity}). Collecting across all paths, we have
\begin{align}
    \frac{\partial \boldsymbol{\tau}}{\partial \mathbf{x}}
= \frac{1}{c}
\begin{bmatrix}
\mathbf{g}_{11}^T \\ \vdots \\ \mathbf{g}_{MN}^T
\end{bmatrix}
 \in \mathbb{R}^{L\times 2}.\label{eq:tau_x}
\end{align}

Because $\mathbf{J}_{\boldsymbol{\eta}\mid\boldsymbol{\alpha}}$ is block diagonal 
with independent delay and Doppler sub-blocks, 
\eqref{eq:chain_rule_FIM} yields
\begin{equation}
\mathbf J_{\mathbf x,\mathbf v}=
\begin{bmatrix}
\frac{1}{c^2}\displaystyle\sum_{(k,\ell)} w^{(\tau)}_{k\ell}\ \mathbf g_{k\ell}\mathbf g_{k\ell}^{\!T}
&
\frac{1}{c}\displaystyle\sum_{(k,\ell)} w^{(\times)}_{k\ell}\ \mathbf g_{k\ell}\mathbf g_{k\ell}^{\!T}
\\[10pt]
\frac{1}{c}\displaystyle\sum_{(k,\ell)} w^{(\times)}_{k\ell}\ \mathbf g_{k\ell}\mathbf g_{k\ell}^{\!T}
&
\displaystyle\sum_{(k,\ell)} w^{(v)}_{k\ell}\ \mathbf g_{k\ell}\mathbf g_{k\ell}^{\!T}
\end{bmatrix}.
\label{eq:JxvFinal}
\end{equation}
When the cross weights vanish ($w^{(\times)}_{k\ell}=0$ for all $(k,\ell)$), the blocks
decouple and the usual geometry-weighted $2\times2$ inverses yield the closed-form CRLBs.
Otherwise, the coupled CRLBs are obtained by $2\times2$ Schur complements of
\eqref{eq:JxvFinal}.

Using the assembled intermediate Fisher matrix in \eqref{eq:JxvFinal}, define the
geometry–weighted sums
\begin{align}
\mathbf G_x  \triangleq  \sum_{(k,\ell)\in L} w^{(\tau)}_{k\ell}\,\mathbf g_{k\ell}\mathbf g_{k\ell}^{\!T},\\
\mathbf G_v  \triangleq  \sum_{(k,\ell)\in L} w^{(v)}_{k\ell}\,\mathbf g_{k\ell}\mathbf g_{k\ell}^{\!T},\\
\mathbf G_{\times}  \triangleq  \sum_{(k,\ell)\in L} w^{(\times)}_{k\ell}\,\mathbf g_{k\ell}\mathbf g_{k\ell}^{\!T}.
\label{eq:III-A-4-Gdefs}
\end{align}
Then the $(\mathbf x,\mathbf v)$–FIM can be written compactly as
\begin{equation}
\mathbf J_{\mathbf x,\mathbf v}  = 
\begin{bmatrix}
\frac{1}{c^2}\,\mathbf G_x & \frac{1}{c}\,\mathbf G_{\times}\\[4pt]
\frac{1}{c}\,\mathbf G_{\times} & \mathbf G_v
\end{bmatrix}.
\label{eq:III-A-4-blockJ}
\end{equation}
The matrices $\mathbf G_x,\mathbf G_v$ are symmetric positive semidefinite (PSD), and
$\mathbf G_{\times}$ is symmetric. Equality $\mathbf G_{\times}=\mathbf 0$ holds when every
path satisfies $\mathrm{Cov}_{k\ell}(t,f_c^2)=0$.

\begin{theorem}[Closed-form CRLBs in multistatic case]
\label{thm:crlb-block}
The $(\mathbf x,\mathbf v)$–FIM in (\ref{eq:III-A-4-blockJ}) can be expressed as:
\begin{equation}
   \mathbf J_{\mathbf x,\mathbf v}
=\begin{bmatrix}
\mathbf A & \mathbf B\\[2pt]
\mathbf B^{\!\top} & \mathbf C
\end{bmatrix},
\quad
\mathbf A=\tfrac{1}{c^{2}}\,\mathbf G_x,\quad
\mathbf B=\tfrac{1}{c}\,\mathbf G_{\times},\quad
\mathbf C=\mathbf G_v, 
\end{equation}
where $\mathbf G_x,\mathbf G_v,\mathbf G_{\times}\in\mathbb{R}^{2\times 2}$ are the position, velocity,
and coupling geometry matrices.
We assume that the network geometry yields full rank so that $\mathbf G_x\succ 0$ and $\mathbf G_v\succ 0$.
Then the CRLB matrix is $\mathbf J_{x,v}^{-1}$ and its blocks admit the closed forms
\begin{equation}
\begin{aligned}\label{eq:CRLB_bi}
\mathrm{CRLB}(\mathbf x)
&= \Big(\mathbf A - \mathbf B\,\mathbf C^{-1}\mathbf B\Big)^{-1}\\
& = \Big(\tfrac{1}{c^2}\big[\mathbf G_x - \mathbf G_{\times}\mathbf G_v^{-1}\mathbf G_{\times}\big]\Big)^{-1},\\[3pt]
\mathrm{CRLB}(\mathbf v)
&= \Big(\mathbf C - \mathbf B^{T}\mathbf A^{-1}\mathbf B\Big)^{-1}\\
&= \Big(\mathbf G_v - c^{2}\,\mathbf G_{\times}\mathbf G_x^{-1}\mathbf G_{\times}\Big)^{-1}.\\[3pt]
\end{aligned}
\end{equation}
Moreover, since $G_{\times}G_v^{-1}G_{\times}^{\!\top}\succeq 0$ and
$G_{\times}^{\!\top}G_x^{-1}G_{\times}\succeq 0$, the coupling strictly reduces
the marginal information in both position and velocity, yielding the bounds
\begin{equation}
\mathrm{CRLB}(x)\succeq c^{2}G_x^{-1},
\qquad
\mathrm{CRLB}(v)\succeq G_v^{-1},
\label{eq:crlb-ineq}
\end{equation}
with equality if and only if $G_{\times}=\mathbf{0}$.
\end{theorem}

\subsubsection{Discussion}\label{sec:discussion}
In the synthetic space–time–frequency design, the CRLBs in \eqref{eq:CRLB_bi} show that position and velocity accuracies are jointly controlled by the synthesized bandwidth, the slow-time span, and the network geometry. Unlike traditional radar/ISAC settings where delay (range) and Doppler largely decouple, frequency hopping couples delay and velocity in this design, so widening the synthesized bandwidth not only sharpens delay resolution but also strengthens velocity information through higher carrier levels. By contrast, extending the slow-time span primarily benefits velocity and provides a more modest gain for position. In practice, bandwidth tends to produce larger overall reductions in both bounds, while additional  CPI yields steadier improvements. Geometry remains pivotal: adding well-placed base stations and diversifying look-angles reduce dilution of precision, with the largest returns from the first few strategically located sites.

\emph{Position (delay) sensitivity:} Delay information grows strongly with the effective fast-time slope of the matched filter, so the position block of the information scales with the squared effective bandwidth (i.e., increases with \(\beta^2\)). Extending the synthetic time \(T_{\rm syn}\) mainly increases the number of pulses \(P\) (a roughly linear gain), whereas increasing the synthesized bandwidth \(B_{\rm syn}\) both enlarges \(\beta\) and raises the carrier-frequency \(\sum f_{c,p}^2\), yielding a markedly larger improvement. 

\emph{Velocity (Doppler) sensitivity:} The per-pulse phase accrued by a moving target is approximately \(\phi_p \approx 2\pi (f_{c,p}/c)\,\mathbf{u}^\top\mathbf{v}\,t_p\), so the velocity information scales with \(\sum t_p^2\,f_{c,p}^2\). Increasing the slow-time span $t_p$ or operating at higher carrier levels \(f_{c,p}\) both boost the velocity block of the information. With \(T_{\rm syn}\) fixed, widening \(B_{\rm syn}\) elevates all \(f_{c,p}^2\) terms uniformly and delivers a pronounced gain in the velocity CRLB. With \(B_{\rm syn}\) fixed, longer apertures still help through the \(t_p^2\) accumulation.

\subsection{CRLBs under Monostatic Setting}\label{sec:monostatic}

This section specializes the multistatic formulation to the monostatic case where each co-located radar pair transmits and only its own receiver observes the echo. The fast/slow-time signal model, orthogonality assumptions, and complex-Gaussian FIM remain as in Section \ref{sec:CRLB_multistatic}, with only the geometric mapping and the path set change. For brevity, we omit repeated derivations and state only the required substitutions and the resulting Fisher information and CRLBs.

Let there be $L$ co-located pairs indexed by $\ell\in\{1,\dots,L\}$ with sensor location $\mathbf{s}_\ell\!\in\!\mathbb{R}^2$. Denote target position $\mathbf{x}\in\mathbb{R}^2$ and velocity $\mathbf{v}\in\mathbb{R}^2$, range $r_\ell=\|\mathbf{x}-\mathbf{s}_\ell\|$, and unit line-of-sight (LOS)
\begin{equation}\label{eq:mono_geomtry}
  \mathbf{u}_\ell \triangleq \frac{\mathbf{x}-\mathbf{s}_\ell}{\|\mathbf{x}-\mathbf{s}_\ell\|}\in\mathbb{R}^2 .  
\end{equation}

For monostatic propagation, we have
\begin{equation}\label{eq:mono-delay}
    \begin{aligned}
        \tau_\ell(\mathbf{x}) &= \frac{2\,r_\ell}{c}, \\
 \nabla_{\mathbf{x}} &= \tau_\ell(\mathbf{x})  \frac{2}{c}\,\mathbf{u}_\ell^{\!T}, \\
f_{D,\ell p}(\mathbf{x},\mathbf{v}) &= \frac{2 f_{c,\ell p}}{c}\,\mathbf{u}_\ell^{\!T}\mathbf{v}, \\
\nabla_{\mathbf{v}} f_{D,\ell p}(\mathbf{x},\mathbf{v}) &= \frac{2 f_{c,\ell p}}{c}\,\mathbf{u}_\ell^{\!T},
    \end{aligned}
\end{equation}
where $f_{c,\ell p}$ is the (possibly frequency-hopped) carrier used by pair $\ell$ on slow-time index $p$. Relative to the multistatic geometry vector $\mathbf{g}_{k\ell}=\mathbf{u}_{t,k}+\mathbf{u}_{r,\ell}$, the monostatic case amounts to the substitution $\mathbf{g}_{\ell}=2\,\mathbf{u}_\ell ,$ and the set of paths reduces from all $(k,\ell)$ to the surviving $\ell$'s. Then the geometry-weighted sums for monostatic case can be written as
\begin{equation}
\left\{
\begin{aligned}
    \bar{\mathbf G}_x &\triangleq 4\sum_{\ell=1}^{L} w^{(\tau)}_\ell\,\mathbf{u}_\ell \mathbf{u}_\ell^{\!T}, \\
\bar{\mathbf G}_v &\triangleq 4\sum_{\ell=1}^{L} w^{(v)}_\ell\,\mathbf{u}_\ell \mathbf{u}_\ell^{\!T}, \\
\bar{\mathbf G}_\times & \triangleq 4\sum_{\ell=1}^{L} w^{(\times)}_\ell\,\mathbf{u}_\ell \mathbf{u}_\ell^{\!T}. 
\end{aligned}
\right..
\label{eq:mono-G}
\end{equation}
Using the chain rule with \eqref{eq:mono-delay}, the joint FI for $[\mathbf{x}^T \mathbf{v}^T]^T$ becomes
\begin{equation}
J_{x,v}^{\text{(mono)}}  = 
\begin{bmatrix}
\frac{1}{c^2}\,\bar{\mathbf G}_x & \frac{1}{c}\,\bar{\mathbf G}_\times\\[4pt]
\frac{1}{c}\,\bar{\mathbf G}_\times & \bar{\mathbf G}_v
\end{bmatrix}.
\label{eq:mono-blockFIM}
\end{equation}
Compared with the multistatic expression, two differences are immediate:
(i) the sum is over the surviving monostatic paths $\ell$ only; and
(ii) each outer product carries a factor $4$ because $\mathbf{g}_\ell\!=\!2\mathbf{u}_\ell$.

\begin{corollary}[CRLBs in monostatic case]
\label{cor:mono}
In the monostatic setting with unit line-of-sight vectors $\mathbf{u}_\ell$,
the bistatic direction reduces to $\mathbf{g}_\ell=2\mathbf{u}_\ell$, and hence substituting (\ref{eq:mono-G}) into Theorem~\ref{thm:crlb-block} gives the monostatic CRLBs. In particular,
if $\bar G_\times=\mathbf{0}$ then
$\mathrm{CRLB}(\mathbf{x})=c^{2}\bar G_x^{-1}$ and
$\mathrm{CRLB}(\mathbf{v})=\bar G_v^{-1}$.
\end{corollary}

\section{Estimation Algorithms}\label{sec:estimators}

In this section, we present parameter-estimation methods for the synthetic ISAC network. We compare two strategies. The first is a fully noncoherent MLE that jointly estimates the target state by maximizing the likelihood of measurements aggregated across BSs. The second is a two-stage information-fusion method that first forms per-BS (per-path) estimations and then fuses them to obtain the final state.

\subsection{Maximum Likelihood Estimation}

Given the received signal in \eqref{eq:received_signal}, we first apply a fast-time matched filter
with respect to the known transmit waveform $s_k(t)$ and sample around the corresponding correlation
peak. This operation compresses the fast-time dimension into a single complex sufficient statistic
per pulse. Stacking these samples across $P$ slow-time pulses, the $(k,\ell)$ path produces a
length-$P$ slow-time observation vector as
\begin{equation}
\mathbf{y}_{k\ell}  =  \alpha_{k\ell}\,\boldsymbol{\phi}_{k\ell}(\mathbf{x},\mathbf{v})  +   \mathbf{w}_{k\ell}, 
\text{ where } \mathbf{w}_{k\ell}\sim\mathcal{CN}(\mathbf{0},\,\sigma_w^2\mathbf{I}_P),
\label{eq:yk-def}
\end{equation}
with slow-time steering vector given by
\begin{equation}
    [\boldsymbol{\phi}_{k\ell}(\mathbf{x},\mathbf{v})]_p
= e^{\!\big(\!-j2\pi f_{c,p}^{k\ell}\,\tau_{k\ell}(\mathbf{x})\big)\,}
  e^{\!\big(\!j2\pi\,t_p\,f_{D,p}^{k\ell}\big),}
\label{eq:steer}
\end{equation}
where $\text{ where } f_{D,p}^{k\ell}  =  \frac{f_{c,p}^{k\ell}}{c}\,\mathbf{g}_{k\ell}^T\mathbf{v}$. After matched
filtering, the fast-time waveform shape is absorbed into the scalar gain $\alpha_{k\ell}$, while the
slow-time structure induced by delay and Doppler is captured by $\boldsymbol{\phi}_{k\ell}$.

Assuming statistical independence across paths and i.i.d.\ circular–Gaussian noise within a path,
the log–likelihood is given by
\begin{equation}
\mathcal{L}(\mathbf{x},\mathbf{v},\boldsymbol{\alpha}) 
= \frac{1}{\sigma_w^2}\sum_{(k,\ell)\in\mathcal{L}}
\big\|\mathbf{y}_{k\ell}-\alpha_{k\ell}\,\boldsymbol{\phi}_{k\ell}(\mathbf{x},\mathbf{v})\big\|_2^2.
\label{eq:nll}
\end{equation}
For fixed $(\mathbf{x},\mathbf{v})$ the log-likelihood is a quadratic function of the complex scalar $\alpha_{k\ell}$. Therefore its MLE solution is available in closed form:
\begin{equation}
\widehat{\alpha}_{k\ell}(\mathbf{x},\mathbf{v})
= \frac{\boldsymbol{\phi}_{k\ell}(\mathbf{x},\mathbf{v})^H\,\mathbf{y}_{k\ell}}
       {\|\boldsymbol{\phi}_{k\ell}(\mathbf{x},\mathbf{v})\|_2^2}.
\label{eq:alphahat}
\end{equation}
Substituting \eqref{eq:alphahat} back into \eqref{eq:nll} concentrates the likelihood by
eliminating $\boldsymbol{\alpha}$, yielding the gain–free objective
\begin{align}
\mathcal{J}(\mathbf{x},\mathbf{v})
&\triangleq -\min_{\boldsymbol{\alpha}} \mathcal{L}(\mathbf{x},\mathbf{v},\boldsymbol{\alpha})
  \stackrel{c}{=}  
\sum_{(k,\ell)\in\mathcal{L}}
\frac{\big|\boldsymbol{\phi}_{k\ell}(\mathbf{x},\mathbf{v})^H\,\mathbf{y}_{k\ell}\big|^2}
     {\|\boldsymbol{\phi}_{k\ell}(\mathbf{x},\mathbf{v})\|_2^2},
\label{eq:concOBJ}
\end{align}
where $\stackrel{c}{=}$ means equality up to an additive constant independent of
$(\mathbf{x},\mathbf{v})$. Therefore, the MLE of $(\mathbf{x},\mathbf{v})$ is given by
\begin{equation}
(\widehat{\mathbf{x}},\widehat{\mathbf{v}})
\in \arg\max_{\mathbf{x},\mathbf{v}} \mathcal{J}(\mathbf{x},\mathbf{v}).
\label{eq:MLE}
\end{equation}
Equation \eqref{eq:concOBJ} shows that the proposed estimator is a sum of normalized
matched filter powers across paths and the complex path gains play no direct role after
concentration. The objective $\mathcal{J}(\mathbf{x},\mathbf{v})$ is generally nonconvex. To curb computational cost, we can employ a coarse to fine (multiresolution) search that evaluates $\mathcal{J}$ on a coarse grid, retains the most promising regions, and iteratively refines them until convergence~\cite{MLE_coarse2fine}.

\subsection{Two-Stage Information Fusion (TSIF)}
\label{subsec:tsif}

\begin{algorithm}[t!]
\caption{Two-Stage Information Fusion (TSIF)}
\label{alg:tsif}
\begin{algorithmic}[1]
\REQUIRE Slow-time data $\{y_k\}_{k=1}^{K}$, pulse times $\{t_p\}$, carriers $\{f_{c,kp}\}$, geometry $\{\mathbf s_{(\cdot),k}\}$, noise variance $\sigma_w^2$, pulse energies $\{E_{s,k}\}$, prior $(\mathbf x^{(0)},\mathbf v^{(0)})$.
\STATE \textbf{Stage A (per-path)}: For each $k$:
\STATE Compute $(\widehat{\tau}_k,\widehat r_k)$ by \eqref{eq:local-ml} on a window around the prior $(\tau_{0,k},r_{0,k})$; refine locally.
\STATE Form $\widetilde{\mathbf J}_k$ using \eqref{eq:perpath-info} and set $\mathbf C_k=\widetilde{\mathbf J}_k^{-1}$.
\STATE \textbf{Stage B (fusion)}: Initialize $(\mathbf x,\mathbf v)\leftarrow(\mathbf x^{(0)},\mathbf v^{(0)})$. Repeat:
\STATE Build residual $\mathbf r_k=\begin{bmatrix}\widehat{\tau}_k\\ \widehat r_k\end{bmatrix}-h_k(\mathbf x,\mathbf v)$ and Jacobian $\mathbf H_k$ from \eqref{eq:hk-def}--\eqref{eq:Hk-Jg}.
\STATE Solve \eqref{eq:GN} for $\delta$ and update $(\mathbf x,\mathbf v)$.
\STATE Stop when $\|\delta\|$ and the cost decrease are below thresholds.
\ENSURE Fused estimates $(\widehat{\mathbf x},\widehat{\mathbf v})$.
\end{algorithmic}
\end{algorithm}

The MLE built from the raw complex slow-time data is asymptotically efficient but can be computationally costly.  A practical alternative is to first form per-path intermediate estimates of delay and radial speed, together with their error covariances, and then fuse these across all paths in a statistically optimal way for the estimations $(\mathbf x,\mathbf v)$.  We term this approach \emph{Two-Stage Information Fusion (TSIF)}.  Under the Gaussian model and high SNR, the per-path estimators are asymptotically normal with covariances given by the Fisher information, and the second-stage fusion uses information-weighted nonlinear least squares, which is consistent and attains the linearized CRLB.

\begin{figure}[!t]
 \centering
 \includegraphics[width=2.5in]{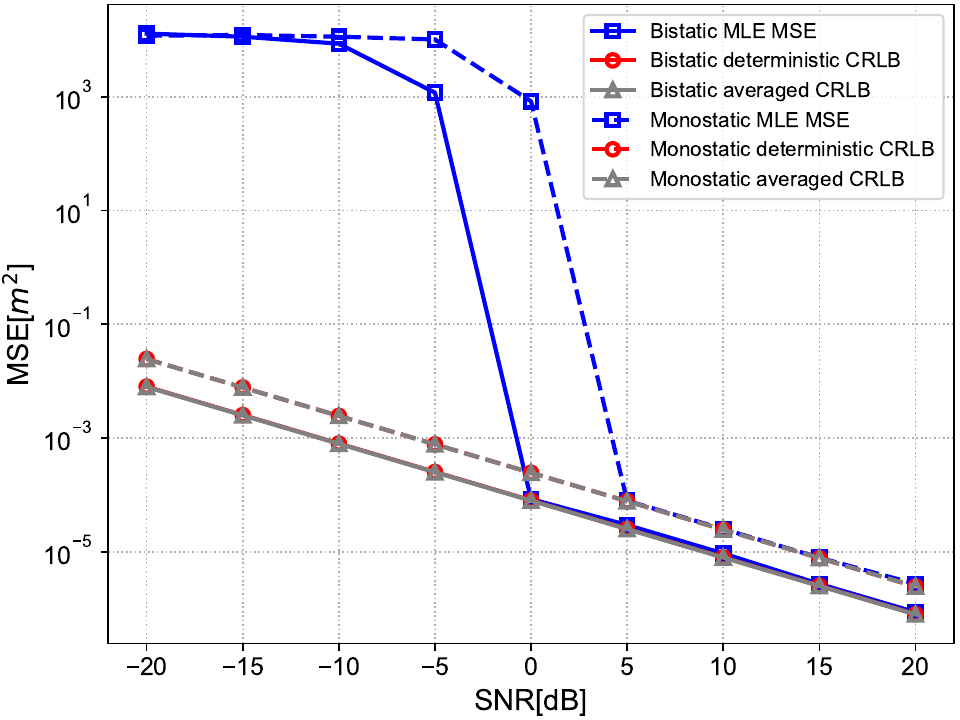}\\
 \caption{MSE of MLE and CRLB of localization versus SNR.}
 \label{fig:MLE_CRLB_pos}
\end{figure}

\begin{figure}[!t]
 \centering
 \includegraphics[width=2.5in]{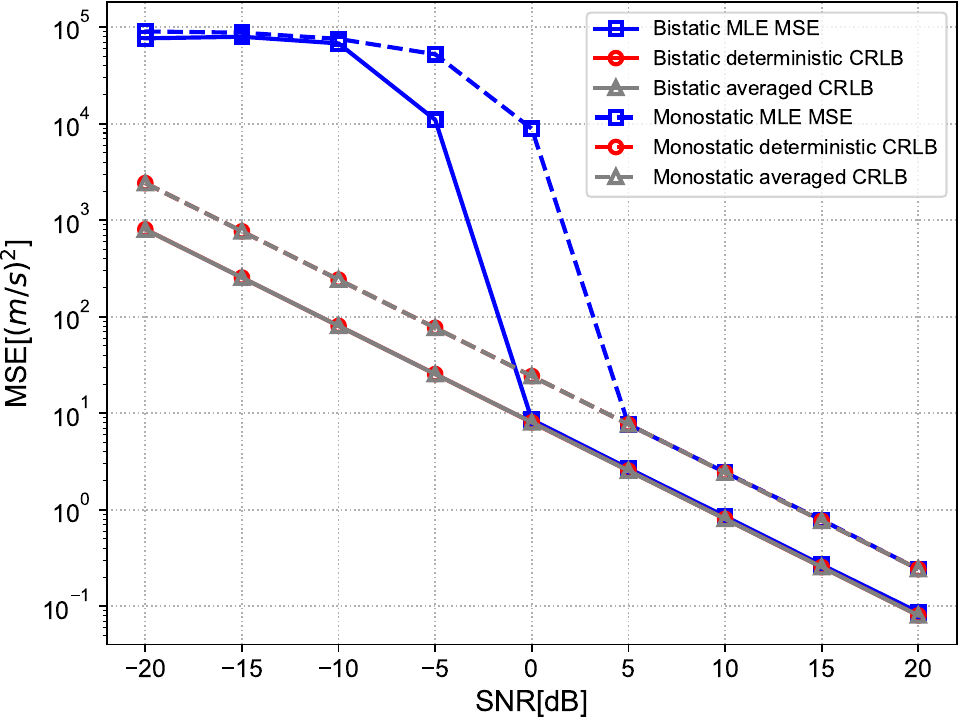}\\
 \caption{MSE of MLE and CRLB of velocity versus SNR.}
 \label{fig:MLE_CRLB_vel}
\end{figure}

\begin{figure*}[!h]
    \centering
    \subfloat[]
    {
        \label{fig:lines_mutual1}
        \includegraphics[width=1.65in]{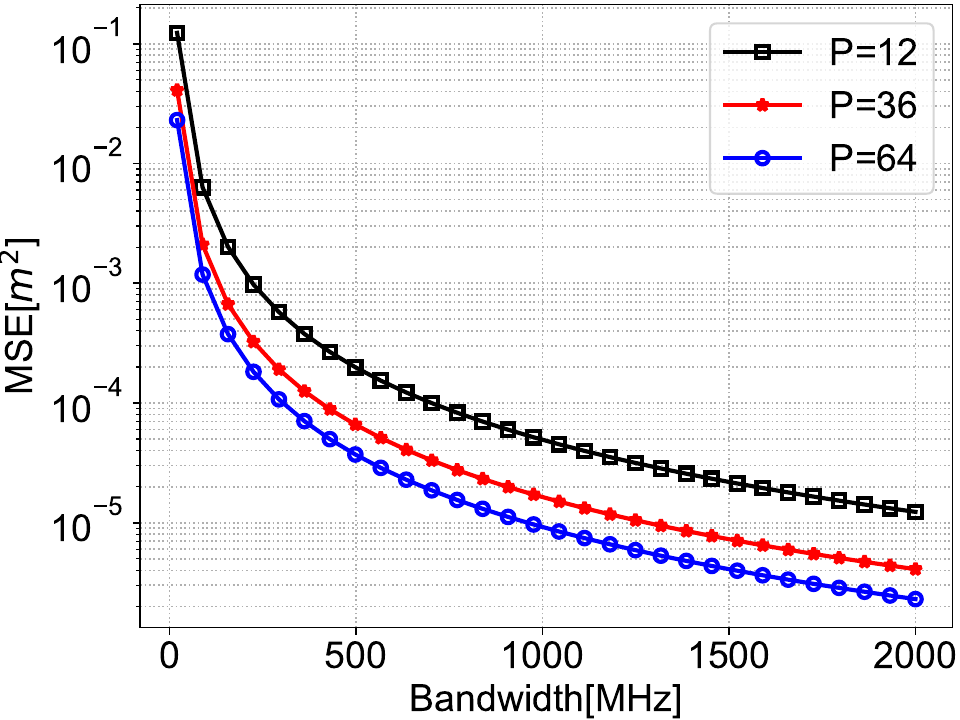}
    }
    \hspace{0.0001\linewidth}
    \subfloat[]
    {
       \label{fig:lines_mutual2}
        \includegraphics[width=1.65in]{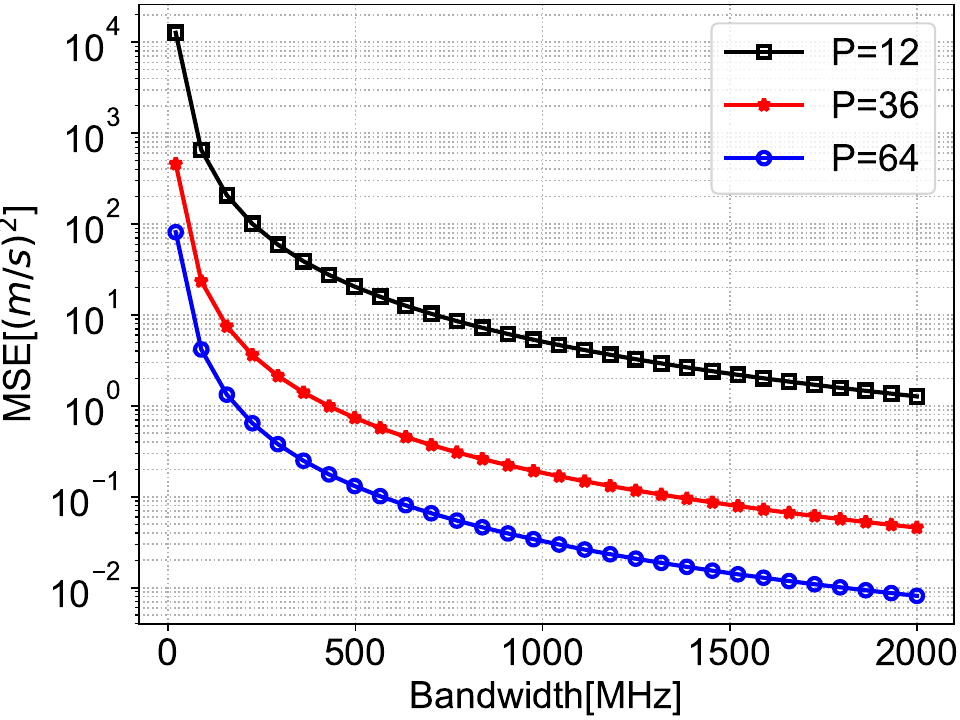}
    }
    \hspace{0.0001\linewidth}
    \subfloat[]
    {
        \label{fig:lines_mutual3}
        \includegraphics[width=1.65in]{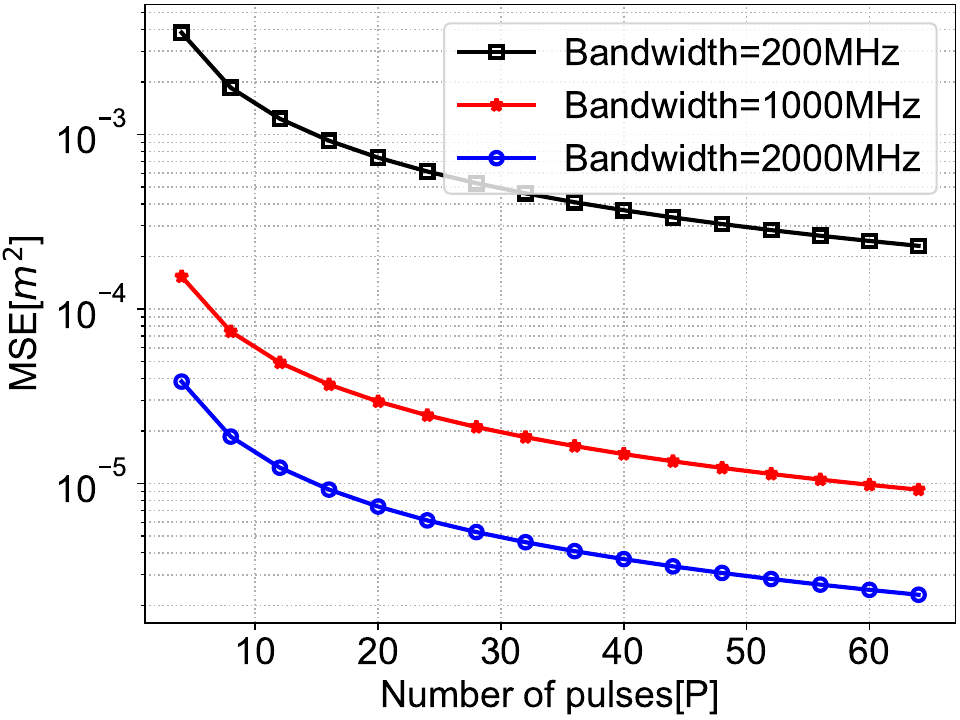}
    }
    \hspace{0.0001\linewidth}
    \subfloat[]
    {
        \label{fig:lines_mutual4}
        \includegraphics[width=1.65in]{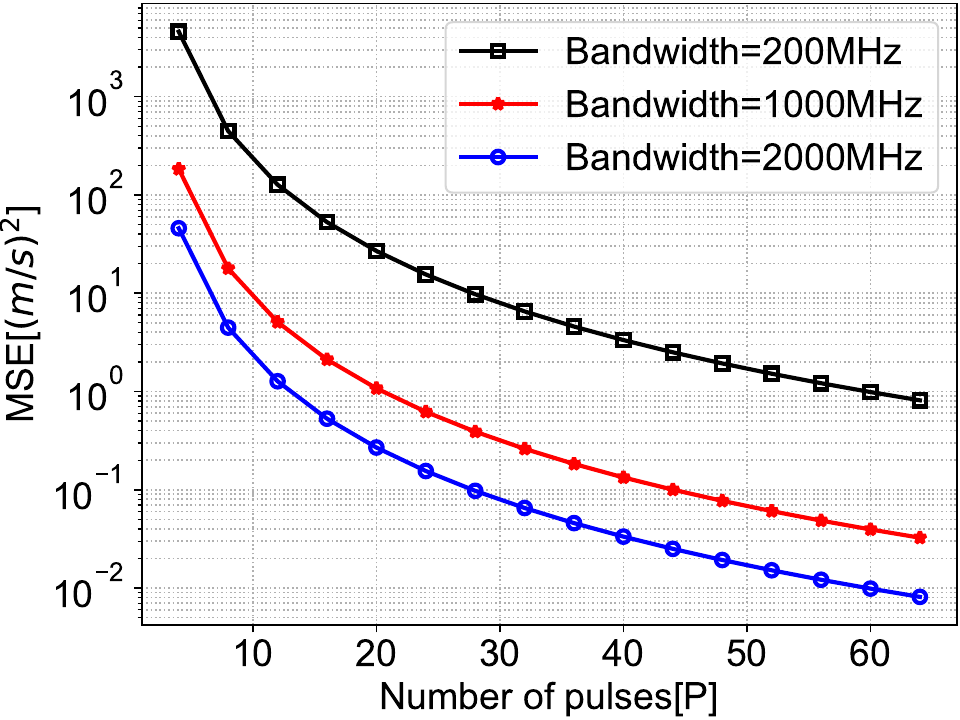}
    }
    
\caption{CRLBs of position and velocity across different synthesized time and bandwidth under both multistatic and monostatic settings: (a) CRLBs of position across different synthesized time and bandwidth under multistatic setting, (b) CRLBs of velocity across different synthesized time and bandwidth under multistatic setting, (c) CRLBs of position across different synthesized time and bandwidth under monostatic setting, (d) CRLBs of velocity across different synthesized time and bandwidth under monostatic setting.}
\label{fig:lines_mutual}
\end{figure*}

\begin{figure*}[!h]
    \centering
    \subfloat[]
    {
        \label{fig:response:aa}
        \includegraphics[width=1.65in]{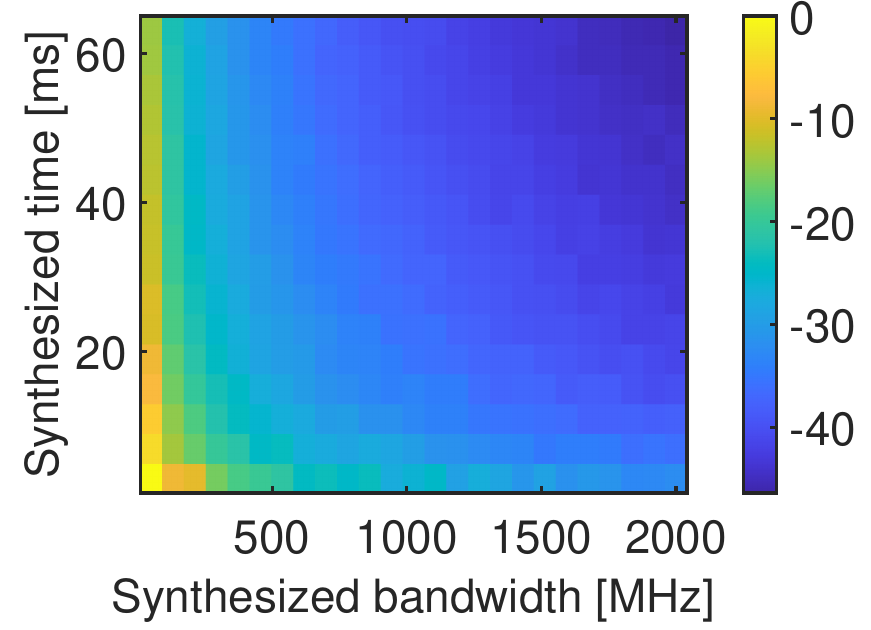}
    }
    \hspace{0.0001\linewidth}
    \subfloat[]
    {
       \label{fig:response2:a}
        \includegraphics[width=1.65in]{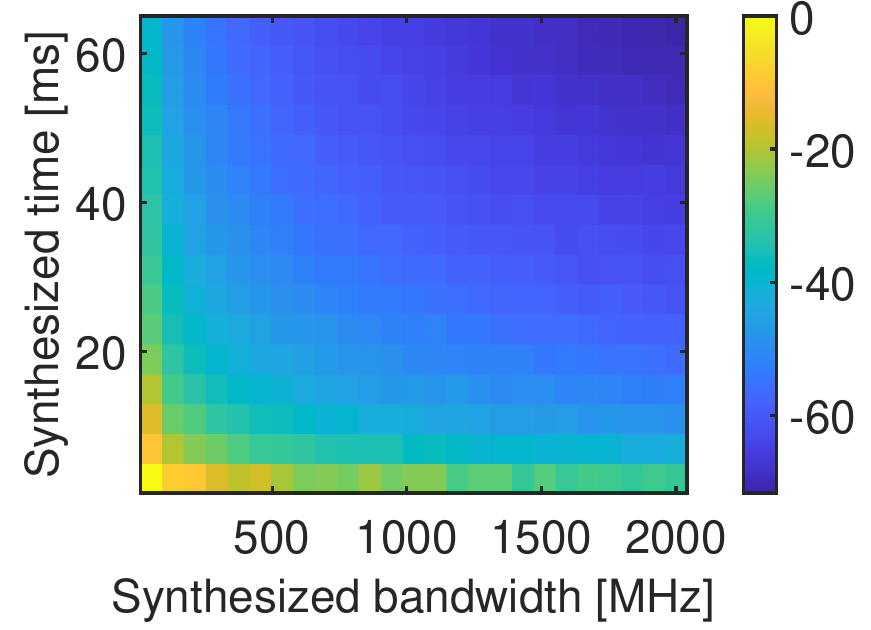}
    }
    \hspace{0.0001\linewidth}
    \subfloat[]
    {
        \label{fig:response:b}
        \includegraphics[width=1.65in]{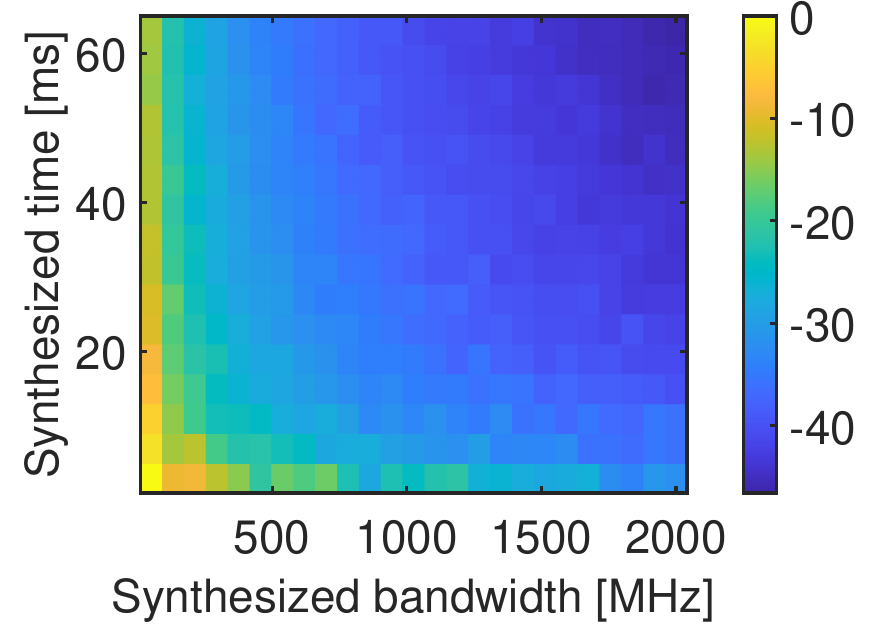}
    }
    \hspace{0.0001\linewidth}
    \subfloat[]
    {
        \label{fig:response:c}
        \includegraphics[width=1.65in]{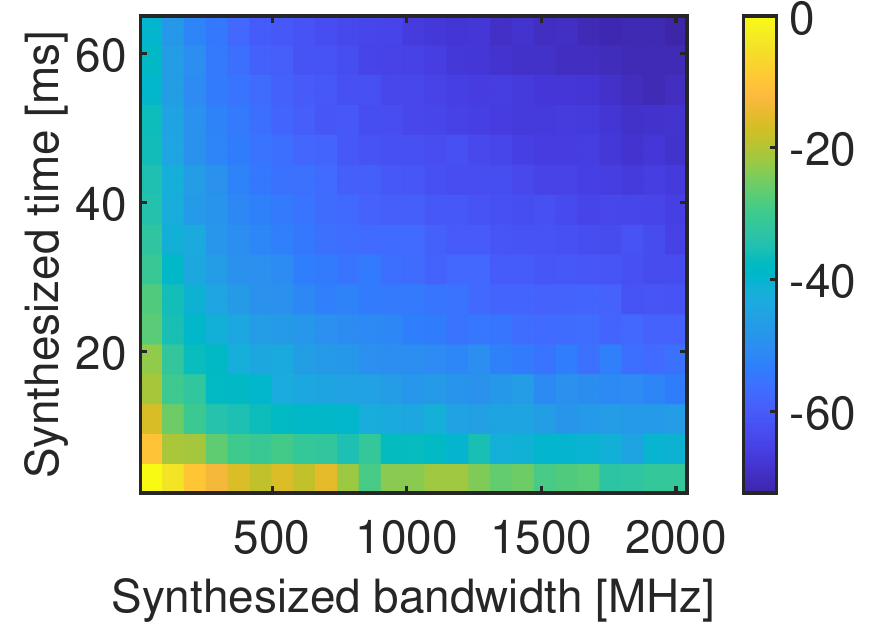}
    }
    
\caption{CRLBs of position and velocity across different synthesized time and bandwidth under both multistatic and monostatic settings: (a) CRLBs of position across different synthesized time and bandwidth under multistatic setting, (b) CRLBs of velocity across different synthesized time and bandwidth under multistatic setting, (c) CRLBs of position across different synthesized time and bandwidth under monostatic setting, (d) CRLBs of velocity across different synthesized time and bandwidth under monostatic setting.}
\label{fig:heatmap_mutual}
\end{figure*}

\subsubsection{Stage A - Per-path Concentrated ML}
For path $k$ (monostatic BS or Tx/Rx pair), let $y_k\in\mathbb{C}^{P}$ be the vector of slow-time samples after waveform separation. The complex baseband received signal 
for the $(k,\ell)$th path at the $p$-th pulse in (\ref{eq:received_signal}) can be rewritten as
\begin{equation}
y_k[p] =\alpha_k\, e^{-j2\pi f_{c,p}^{k}\,\tau_k}\,
               e^{j2\pi (f_{c,p}^{k}/c)\,r_k\, t_p}  +  w_k[p],
\label{eq:slowtime-model}
\end{equation}
where $\alpha_k$ denotes the complex amplitude, $\tau_k$ denotes the delay, and $r_k \triangleq \mathbf g_k(\mathbf x)^{\!T}\mathbf v$ is the radial velocity. $\{w_k[p]\}$ is circular complex Gaussian with variance $\sigma_w^2$.  Define the per-path steering vector as
\begin{equation}
\begin{aligned}
   \phi_k(\tau,r) &\triangleq  \mathbf a_k(\tau)\odot \mathbf d_k(r),
\\
[\mathbf a_k(\tau)]_p &= e^{-j2\pi f_{c,p}^k\tau},  \\
[\mathbf d_k(r)]_p& = e^{\,j2\pi (f_{c,p}^k/c)\, r\, t_p}. 
\end{aligned}
\end{equation}
Concentrating out $\alpha_k$ yields the generalized likelihood ratio test (GLRT) objective
\begin{equation}
\widehat{\tau}_k,\widehat{r}_k
 \in \arg\max_{\tau,r} 
\Lambda_k(\tau,r)
 \triangleq 
\frac{\big|\phi_k(\tau,r)^{\!H} y_k\big|^{2}}{\|\phi_k(\tau,r)\|_2^{2}} .
\label{eq:local-ml}
\end{equation}
Practical estimations uses a windowed grid. Unlike Section~\ref{sec:CRLB}, where information is expressed for the true velocity vector $\mathbf{v}$, here we reparameterize the Doppler by the per-path radial speed $r_k = \mathbf{g}_k(\mathbf x)^{\!T}\mathbf v$ yields (\ref{eq:perpath-info}), in which the geometry factors $\mathbf{g}_k$ are absorbed by the Jacobian, leaving only slow-time variance and covariance terms:
\begin{equation}
\begin{aligned}
\widetilde{\mathbf J}_k & = 
\begin{bmatrix}
A_k & -B_k\\[2pt]
-B_k & D_k
\end{bmatrix},
\\
A_k &= \frac{8\pi^2}{\sigma_w^2}\,|\tilde{\alpha}_k|^2 E_{s,k}\, P_k \mathrm{Var}_p\!\big(f_{c,p}^k\big),\\
B_k &= \frac{8\pi^2}{\sigma_w^2}\,|\tilde{\alpha}_k|^2 E_{s,k}\, \frac{P_k}{c} \mathrm{Cov}_p\!\big(t_p,{f_{c,p}^k}^2\big),\\
D_k &= \frac{8\pi^2}{\sigma_w^2}\,|\tilde{\alpha}_k|^2 E_{s,k}\, \frac{P_k}{c^{2}} \mathrm{Var}_p\!\big(t_p f_{c,p}^k\big),
\end{aligned}
\label{eq:perpath-info}
\end{equation}
where $|\tilde{\alpha}_k|$ denotes the complex amplitude, $E_{s,k}$ the per-pulse energy, and the variances/covariances are taken across $p=0,\dots,P_k-1$ (centering $t_p$ and $f_{c,kp}$ if desired).  The per-path covariance is then
\begin{equation}
\mathbf C_k  =  \widetilde{\mathbf J}_k^{-1}
 = 
\begin{bmatrix}
\sigma^2_{\tau,k} & \sigma_{\tau r,k}\\
\sigma_{\tau r,k} & \sigma^2_{r,k}
\end{bmatrix}.
\label{eq:perpath-cov}
\end{equation}

\subsubsection{Stage B - Weighted Gauss--Newton fusion}
We stack the per-path estimates $\mathbf z_k \triangleq [\widehat{\tau}_k, \widehat{r}_k]^{T}$ and model them via the following deterministic mapping
\begin{equation}
h_k(\mathbf x,\mathbf v)
 = 
\begin{bmatrix}
\tau_k(\mathbf x)\\[1pt]
r_k(\mathbf x,\mathbf v)
\end{bmatrix}
=
\begin{bmatrix}
\frac{1}{c}\big(R_{t,k}(\mathbf x)+R_{r,k}(\mathbf x)\big)\\[2pt]
\mathbf g_k(\mathbf x)^{T}\mathbf v
\end{bmatrix},
\label{eq:hk-def}
\end{equation}
where $R_{(\cdot),k}(\mathbf x)$ is the one-way radar-to-target range and
\begin{equation}
\mathbf g_k(\mathbf x)=
\begin{cases}
2\,\dfrac{\mathbf x-\mathbf s_k}{\|\mathbf x-\mathbf s_k\|}, & \text{monostatic},\\[10pt]
\dfrac{\mathbf x-\mathbf t_{k}}{\|\mathbf x-\mathbf t_{k}\|}+\dfrac{\mathbf x-\mathbf r_{k}}{\|\mathbf x-\mathbf r_{k}\|}, & \text{multistatic}.
\end{cases}
\label{eq:gk-def}
\end{equation}
Linearizing $h_k$ at a current iteration $(\mathbf x,\mathbf v)$ gives the Jacobian
\begin{equation}
\begin{aligned}
    \mathbf H_k
&=
\begin{bmatrix}
\frac{1}{c}\,\mathbf g_k(\mathbf x)^{T} & \mathbf 0_{1\times 2}\\[4pt]
\big(\mathbf J_{g,k}(\mathbf x)\,\mathbf v\big)^{T} & \mathbf g_k(\mathbf x)^{T}
\end{bmatrix},
\\
\mathbf J_{g,k}(\mathbf x)
&=
\begin{cases}
\dfrac{2}{\|\mathbf x-\mathbf s_k\|}\!\left(\mathbf I-\hat{\mathbf u}_k\hat{\mathbf u}_k^{T}\right), & \text{monostatic},\\[10pt]
\dfrac{\mathbf I-\hat{\mathbf u}_{t,k}\hat{\mathbf u}_{t,k}^{T}}{R_{t,k}}+\dfrac{\mathbf I-\hat{\mathbf u}_{r,k}\hat{\mathbf u}_{r,k}^{T}}{R_{r,k}}, & \text{bistatic},
\end{cases}
\end{aligned}
\label{eq:Hk-Jg}
\end{equation}
where $\hat{\mathbf u}=\frac{\mathbf x-\mathbf s}{\|\mathbf x-\mathbf s\|}$ denotes the radar to target unit vectors.  With weights $\mathbf W_k=\mathbf C_k^{-1}$, one Gauss--Newton step for $\boldsymbol\theta\!\triangleq[\mathbf x^{T},\mathbf v^{T}]^{T}$ solves the normal equations
\begin{equation}
\Big(\sum_{k}\mathbf H_k^{T}\mathbf W_k \mathbf H_k\Big)\,\delta
 = 
\sum_{k}\mathbf H_k^{T}\mathbf W_k\,\big(\mathbf z_k-h_k(\mathbf x,\mathbf v)\big).
\label{eq:GN}
\end{equation}
After solving for $\delta$, update the estimate as $\boldsymbol{\theta}\leftarrow \boldsymbol{\theta}+\delta$ and iterate until convergence. 


\section{Simulation Results}\label{sec:results}

\begin{figure*}[!t]
    \centering
    \subfloat[]
    {
        \label{fig:heatmap_BS_pos1}
        \includegraphics[width=2.0in]{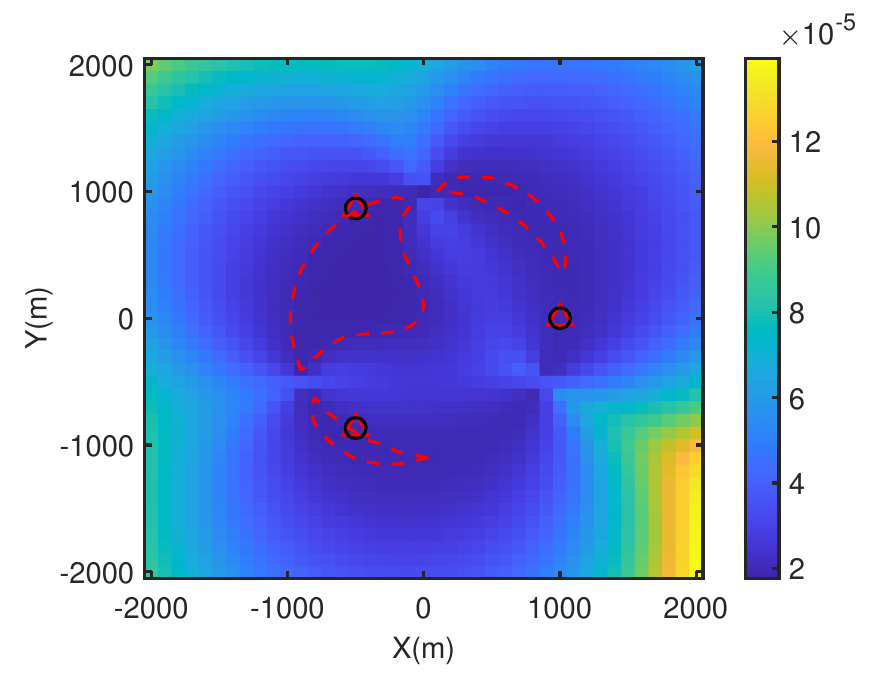}
    }
    \hspace{0.0001\linewidth}
    \subfloat[]
    {
       \label{fig:heatmap_BS_pos2}
        \includegraphics[width=2.0in]{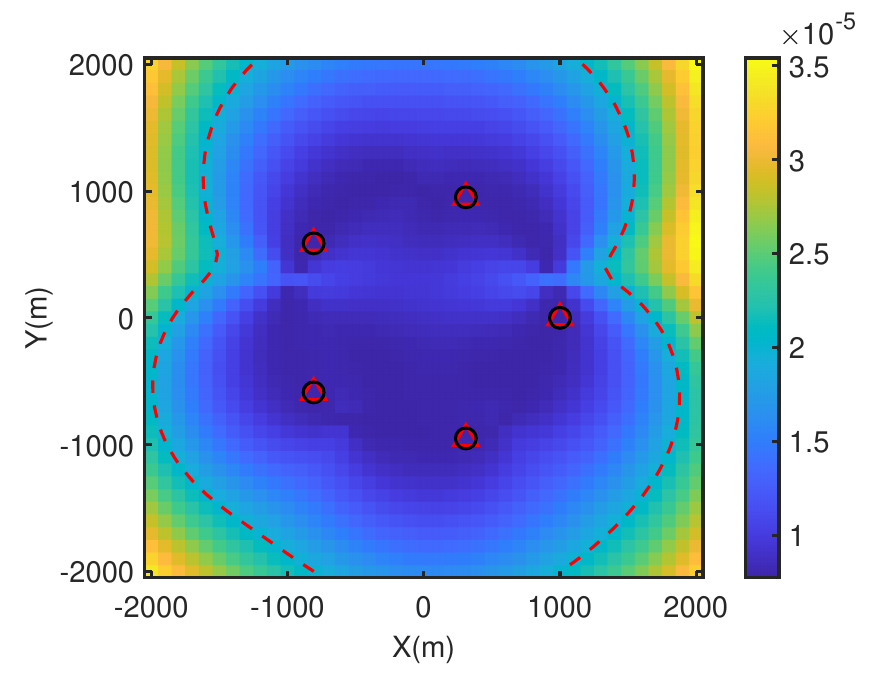}
    }
    \hspace{0.0001\linewidth}
    \subfloat[]
    {
        \label{fig:heatmap_BS_pos3}
        \includegraphics[width=2.0in]{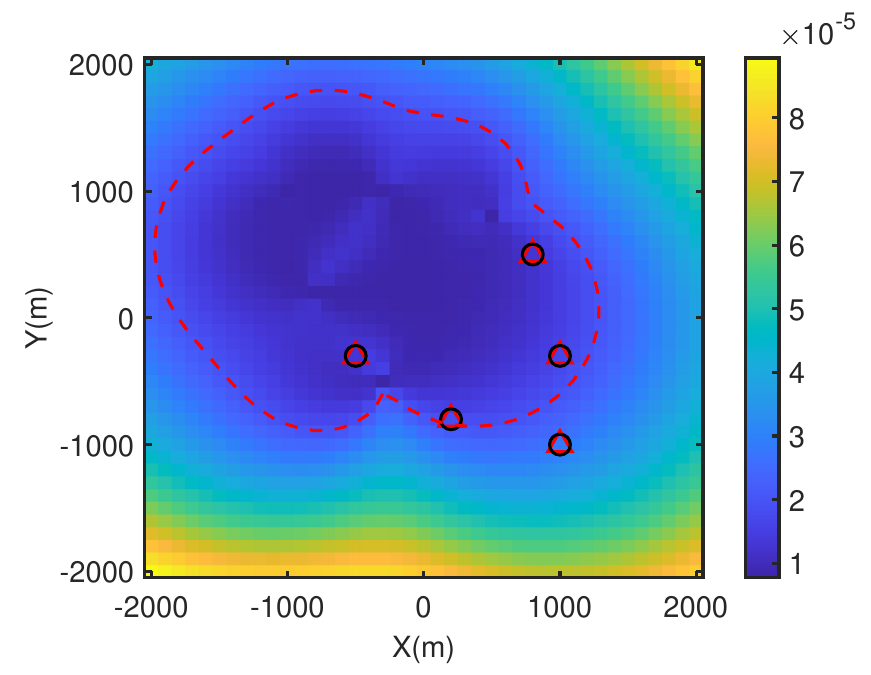}
    }
    
\caption{CRLBs of position with different number of BSs and locations in monostatic setting: (a) CRLBs of position with 3 BSs in uniform circle, (b) CRLBs of position with 5 BSs in uniform circle, (c) CRLBs of position with 5 BSs in random locations.}
\label{fig:heatmap_BS_pos}
\end{figure*}


In this study, we explore a synthesized ISAC system characterized by
specific parameters. Unless otherwise specified, the carrier is $f_0=28~\mathrm{GHz}$ and the per–pulse carriers hop uniformly over a synthesized span of $\Delta f_{\max}=2~\mathrm{GHz}$ around $f_0$. The slow–time aperture consists of $P=12$ pulses with pulse–repetition interval $\mathrm{PRI}=1~\mathrm{ms}$. For geometry, we primarily consider a multistatic network with $M=N=3$ transmitters and receivers placed on concentric rings of radius $1000\mathrm{m}$; when a monostatic configuration is used, it comprises $L=5$ colocated Tx/Rx BSs uniformly placed on a circle of radius $1000\mathrm{m}$. The ground–truth state is $\mathbf{x}=[300,\,200]\,\mathrm{m}$ and $\mathbf{v}=[20,\,15]\,\mathrm{m/s}$. We normalize the per-pulse fast-time energy to unit energy and set $E_s=1$. We report the Monte Carlo MSE $\mathbb{E}\!\left[\|\hat{\mathbf{x}}-\mathbf{x}\|_2^2\right]$ and $\mathbb{E}\!\left[\|\hat{\mathbf{v}}-\mathbf{v}\|_2^2\right]$ as functions of SNR and overlay the derived CRLB. Unless otherwise stated, per SNR is averaged over $10^3$ independent noise realizations with a fixed random seed to ensure reproducibility. Note that this paper focuses on the sensing gains enabled by the proposed synthesis strategy. The communication functionality is identical to that of a conventional frequency-hopping communication system. Therefore, for clarity and due to page limitations, we do not further analyze the communication performance.

We first assess the MLE under the proposed model by computing the CRLBs and the empirical MSEs for both multistatic and monostatic configurations. Note that the CRLB analysis in Section~\ref{sec:CRLB} is carried out under a deterministic waveform model, where the transmit signal is assumed to be perfectly known at the sensing receiver (e.g., a dedicated radar waveform or a known pilot/preamble). In a genuinely ISAC scenario, however, the transmit signal is randomized to convey communication data. To quantify the impact of this randomness, we also evaluate a data-averaged CRLB corresponding to the unconditional FIM. Specifically, for a given system configuration and parameter vector $\boldsymbol{\theta}$, we draw multiple independent realizations of the communication data symbols according to the employed modulation format (e.g., 16-QAM on the OFDM subcarriers), generate the corresponding time-domain transmit waveform, and compute the FIM for each realization. The data-averaged FIM is then obtained via Monte Carlo averaging over these realizations. In the simulations, we use $N_{\mathrm{sub}}=1024$ subcarriers with a subcarrier spacing of $1.62$~kHz and 16-QAM modulation, and we average over $5\times10^3$ independent data realizations. The resulting data-averaged CRLB is compared with the deterministic-waveform CRLB computed using an effective bandwidth of $48$~MHz, which corresponds to the empirically averaged effective bandwidth of the OFDM waveform. Figs.~\ref{fig:MLE_CRLB_pos} and \ref{fig:MLE_CRLB_vel} report the MSE versus SNR for position and velocity, respectively, together with the corresponding CRLBs for both the deterministic and data-averaged settings. We observe that the data-averaged CRLB induced by communication data randomness closely aligns with the deterministic CRLB. Therefore, in the subsequent results we use the deterministic CRLB as the performance baseline. As expected, at high SNR the MLE is asymptotically unbiased and its MSE approaches the CRLB. At low SNR, sidelobes and noise increasingly corrupt performance, rendering the CRLB non-tight.
In our setup the multistatic system achieves a lower CRLB than the monostatic system. The reason is purely informational: with \(M=3\) (transmitters) and \(N=3\) (receivers), the multistatic array provides \(MN=9\) independent Tx–Rx paths, whereas the monostatic array with \(L=3\) co-located pairs yields only \(L=3\) self-echo paths. Because the Fisher information adds across independent paths, fewer paths translate into a smaller information matrix and hence a larger CRLB. This gap would narrow if the monostatic system employed more pairs or a more favorable geometry/waveform design.

We then examine the joint effect of synthesized bandwidth and time on the estimation accuracy. We first fix the synthesized slow-time length to \(P\in\{12,36,64\}\) pulses and sweep the synthesized bandwidth \(B_{\rm syn}\in[50,2000]\)~MHz. We then fix \(B_{\rm syn}\in\{200,1000,2000\}\)~MHz and sweep the slow-time length \(P\in[4,64]\) pulses. The resulting CRLB trends in Fig.~\ref{fig:lines_mutual} show that, with either \(P\) or \(B_{\rm syn}\) fixed, increasing the other parameter monotonically decreases the CRLBs for both position and velocity. To further visualize the joint trade-space, we span \((B_{\rm syn},T_{\rm syn})\in[50,2000]\ \text{MHz}\times[4,64]\ \text{ms}\) for each pair, and compute CRLBs under both multistatic and monostatic configurations. Within each configuration we normalize the CRLBs for fair visual comparison in Fig.~\ref{fig:heatmap_mutual}. Fig.~\ref{fig:heatmap_mutual} shows that, holding one axis fixed and increasing the other, the CRLBs decrease monotonically:
(i) at fixed \(B_{\rm syn}\), enlarging \(T_{\rm syn}\) reduces the CRLBs, with a modest gain for position and a larger gain for velocity; 
(ii) at fixed \(T_{\rm syn}\), widening \(B_{\rm syn}\) yields a substantial reduction for both position and velocity. This observation is aligned with the discussion in Section~\ref{sec:discussion}, it gives us the insight that when bandwidth and time are in competition, prioritizing synthesized bandwidth generally offers greater overall payoff as it quadratically enhances fast-time delay information and simultaneously strengthens Doppler sensitivity through the \(f_{c,p}^2\) factor, whereas extending time provides linear gains. This guideline holds for both multistatic and monostatic ISAC schemes.

We also evaluate how the placement and number of BSs affect the CRLBs for position and velocity. Without loss of generality, we focus on the monostatic configuration, which is common in practice. Two BS counts are considered 3 and 5 over a \(4\,\text{km}\times4\,\text{km}\) area \([{-}2000,2000]\text{ m}\times[{-}2000,2000]\text{ m}\). We compare a symmetric layout with BSs evenly placed on a circle of radius \(1000\) m, shown in Fig.~\ref{fig:heatmap_BS_pos1} where red triangles denote transmitters and black circles denote co-located receivers, against a random layout, shown in Fig.~\ref{fig:heatmap_BS_pos3}, in which the BSs are not evenly distributed. For each location on a grid covering the area, we compute the CRLBs for position and velocity under each layout. The resulting heatmaps for position are shown in Fig.~\ref{fig:heatmap_BS_pos}. Note that the velocity estimation performance is not explicitly shown, since it closely follows the position estimation results. This similarity arises because both bounds share the same geometric weighting structure, as discussed in Section~\ref{sec:CRLB}, and is therefore omitted for clarity. For visualization and coverage comparison, we set thresholds of \(2\times10^{-5}\,\text{m}^2\) for position CRLB, and overlay contours delineating regions where the CRLBs fall below these thresholds.
Increasing the number of BSs from \(3\) to \(5\) substantially expands the below-threshold region in Fig.~\ref{fig:heatmap_BS_pos}, indicating a marked improvement in the position accuracy, and the same trend is evident for velocity. When the layout departs from symmetry as shown in Fig.~\ref{fig:heatmap_BS_pos3}, pockets of good accuracy remain, while the overall coverage shrinks relative to the circular deployment. When the viewing angle of the target is very restricted, there is a remarkable degradation of performance. To sum up, the results in Figs.~\ref{fig:lines_mutual}–\ref{fig:heatmap_BS_pos} demonstrate that the proposed space–time–frequency synthetic ISAC network yields substantially improved position and velocity accuracy as the synthesized bandwidth, CPI, and BS geometry are enhanced, significantly outperforming typical unsynthesized configurations.

\begin{figure}[!t]
 \centering
 \includegraphics[width=2.5in]{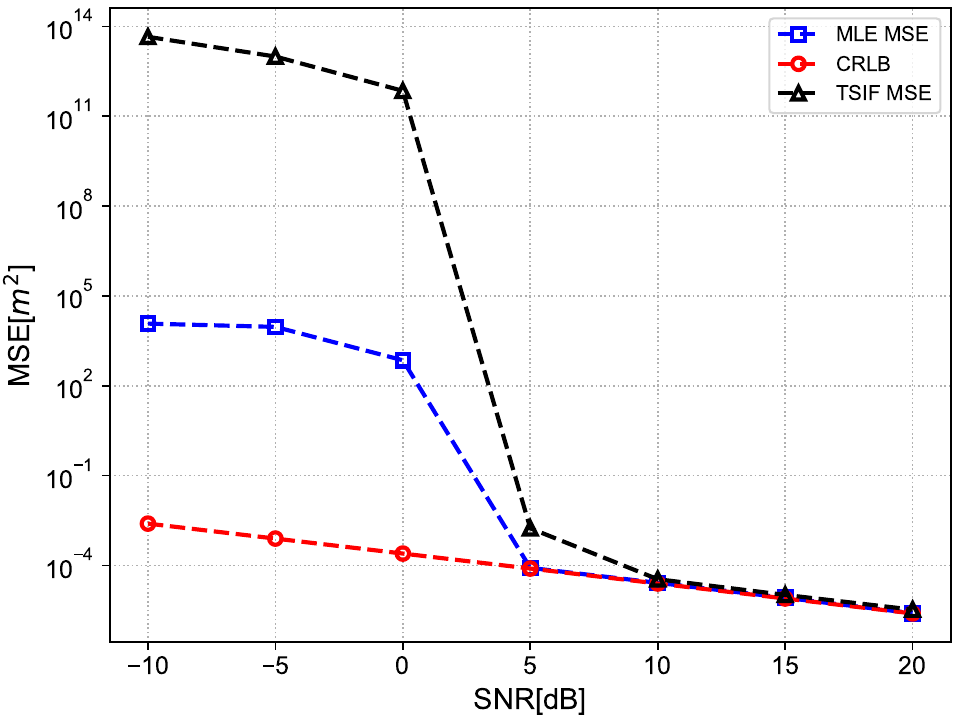}\\
 \caption{MSE of MLE and TSIF of localization versus SNR.}
 \label{fig:MLE_TSIF_pos}
\end{figure}

\begin{figure}[!t]
 \centering
 \includegraphics[width=2.5in]{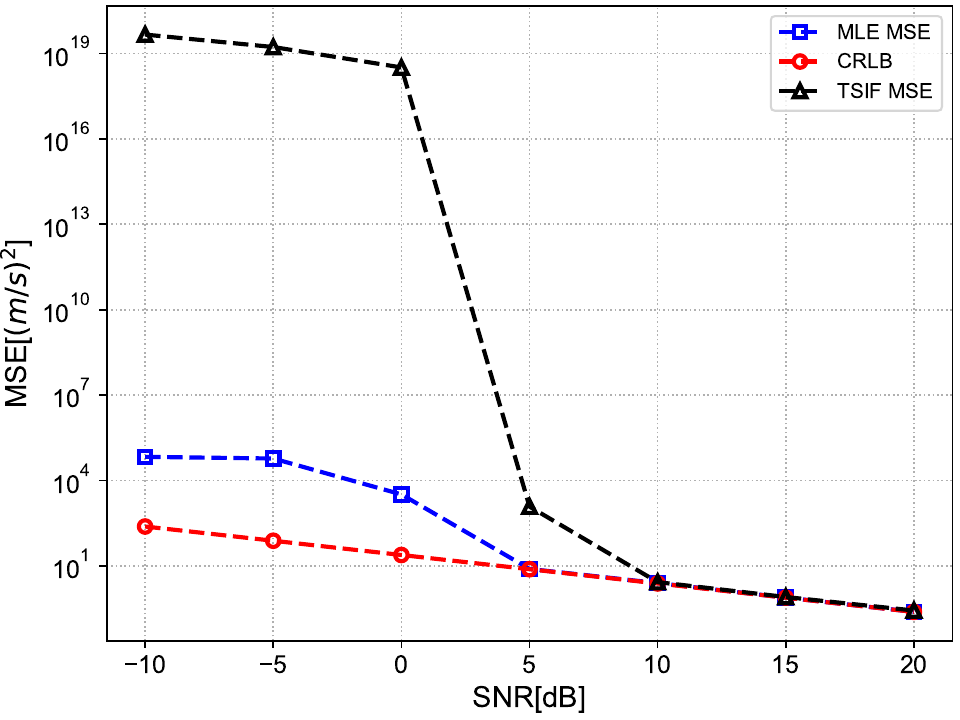}\\
 \caption{MSE of MLE and TSIF of velocity versus SNR.}
 \label{fig:MLE_TSIF_vel}
\end{figure}

Finally, we compare the estimation performance of TSIF and MLE. We consider a monostatic network with 5 base stations placed evenly on a circle of radius 1000\,m, and we evaluate position and velocity accuracy as SNR varies, averaging over independent noise realizations. The results in Figs.~\ref{fig:MLE_TSIF_pos} and \ref{fig:MLE_TSIF_vel} show a clear trend. At high SNR, TSIF closely tracks MLE for both position and velocity, and MLE approaches the corresponding CRLBs. As SNR decreases to the moderate regime, TSIF begins to exhibit a gap to the MLE. This behavior is consistent with the structure of the two methods. TSIF relies on per-path delay and radial speed estimates from Stage A together with their covariances, then fuse them by solving a weighted nonlinear least-squares problem via the Gauss-Newton algorithm in Stage B. When SNR is low, the per-path estimates become noisy and occasionally suffer outages, their error distributions deviate from Gaussian distribution, and the estimated covariances can be misspecified, which amplifies error propagation during fusion. In contrast, the concentrated MLE operates directly on the complex slow time data and jointly fits the global position and velocity, which preserves more of the informative structure in the measurements and is less sensitive to early decision errors. The comparison confirms our central finding. Fully synthesized processing of network measurements should be preferred whenever reliability is critical, while TSIF offers a useful lower complexity alternative in moderate to high SNR conditions where its performance remains close to that of MLE.

\section{Conclusion}\label{sec:conclusion}
This paper established a synthetic ISAC framework that combines observations across space, time, and frequency to overcome limits in aperture, bandwidth, and coherent observation time. We built a unified signal model for multistatic and monostatic settings, derived the concentrated CRLBs for joint position and velocity, and revealed a hopping induced coupling between delay and velocity that clarifies the role of observation scheduling. Experiments showed that concentrated MLE approaches the bound at high SNR, whereas direct fusion of per-BS estimates is fragile at low SNR. These findings lead to a clear design guidance for dense sensing networks: favor fully synthesized processing on network-wide measurements, prioritize synthesized bandwidth before added  CPI, enlarge geometric aperture with diverse base station placement, and select hopping schedules that decorrelate time and carrier usage. The framework provides a basis for upgrading existing communication infrastructure into sensing systems.

\begin{appendices}

\section{Detailed Derivations of FIM (\ref{eq:Jaa})–(\ref{eq:Jtau_alpha_zero})(\ref{eq:delay_velocity_cross_term})}\label{appendix:FIM}

\subsection*{A.1   Amplitude–amplitude block \texorpdfstring{$\mathbf{J}_{\alpha\alpha}$}{J\_aa} (Eq.\ (\ref{eq:Jaa}))}
Given \eqref{eq:FIM_1}\eqref{eq:FIM_2} and $\|(d\odot a)\|_2^2=\sum_p |d_p|^2|a_p|^2=P$, we have
\begin{align}
J_{\Re\alpha,\Re\alpha}
&=\frac{2}{\sigma_w^2}\,\|(d\odot a)\otimes s\|_2^2
=\frac{2}{\sigma_w^2}\,\|(d\odot a)\|_2^2 \,\|s\|_2^2
=\frac{2}{\sigma_w^2}\, P E_s,\\
J_{\Im\alpha,\Im\alpha}
&=\frac{2}{\sigma_w^2}\,\|j(d\odot a)\otimes s\|_2^2
=\frac{2}{\sigma_w^2}\, P E_s,\\
J_{\Re\alpha,\Im\alpha}&=0.
\end{align}
Then given \eqref{eq:FIM_complex}, $J_{\alpha\alpha}$ can be calculated as
\begin{equation}
   \mathbf{J}_{\alpha\alpha}
=\frac{2PE_s}{\sigma_w^2}\, I_2. 
\end{equation}

\subsection*{A.2   Delay–delay element \texorpdfstring{$J_{\tau\tau}$}{J\_tt} (Eq.\ (\ref{eq:Jtautau}))}
Split \eqref{eq:FIM_3} into two orthogonal pieces:
\[
u_1=(d\odot a)\otimes \partial_\tau s,
\qquad
u_2=(d\odot\partial_\tau a)\otimes s.
\]
Then
\begin{align}
\|u_1\|_2^2
&=\|(d\odot a)\|_2^2\,\|\partial_\tau s\|_2^2
=P\cdot (4\pi^2\beta^2 E_s),\\
\|u_2\|_2^2
&=\sum_p |d_p|^2\,|\partial_\tau a_p|^2\,\|s\|_2^2
=\sum_p (2\pi f_{c,p})^2\,E_s
=4\pi^2 F_2\,E_s.
\end{align}
The cross inner product satisfies
\begin{equation}
    \Re\big\{\langle u_1,u_2\rangle\big\}
=\Re\Big\{\langle (d\odot a),(d\odot\partial_\tau a)\rangle\,
\langle \partial_\tau s, s\rangle\Big\}
=0,
\end{equation}
because $\Re\{\langle \partial_\tau s, s\rangle\}
=\tfrac{1}{2}\partial_\tau \|s(\tau)\|_2^2
=\tfrac{1}{2}\partial_\tau E_s=0$ (the matched-filter snapshot energy does not depend on $\tau$).
Therefore,
\begin{equation}
\left\|\frac{\partial \mu}{\partial \tau}\right\|_2^2
=|\alpha|^2\big(\|u_1\|_2^2+\|u_2\|_2^2\big)
=|\alpha|^2\,4\pi^2 E_s\,( \beta^2 P + F_2),
\end{equation}
and \eqref{eq:FIM_complex} yields
\begin{equation}
J_{\tau\tau}
=\frac{2}{\sigma_w^2}\left\|\frac{\partial \mu}{\partial \tau}\right\|_2^2
=\frac{8\pi^2}{\sigma_w^2}\,|\alpha|^2 E_s\big(\beta^2 P + F_2\big).
\end{equation}

\subsection*{A.3   Velocity–velocity block \texorpdfstring{$\mathbf{J}_{\mathbf{vv}}$}{J\_vv} (Eq.\ (\ref{eq:Jff}))}
From \eqref{eq:FIM_4}, using $\| (\partial_v d\odot a)\otimes s \|_2^2=\|s\|_2^2\sum_p\| \partial_v d_p\|_2^2$ and
$\partial_v d_p = j2\pi t_p \tfrac{f_{c,p}}{c} d_p\, \mathbf{g}^T$,
\begin{align}
    \sum_p\|\partial_v d_p\|_2^2
=\sum_p (2\pi)^2 t_p^2 \left(\frac{f_{c,p}}{c}\right)^2 \|\mathbf{g}\|_2^2
\\
\left\|\frac{\partial \mu}{\partial v}\right\|_2^2
=|\alpha|^2 E_s\,(2\pi)^2 \sum_p \left(\frac{f_{c,p}}{c}\right)^2 t_p^2\, \mathbf{g} \mathbf{g}^T.
\end{align}
Thus
\begin{equation}
\mathbf{J}_{\mathbf{v}\mathbf{v}}
=\frac{2}{\sigma_w^2}\left\|\frac{\partial \mu}{\partial v}\right\|_2^2
=\frac{8\pi^2}{\sigma_w^2}\,|\alpha|^2 E_s\,
\sum_p \left(\frac{f_{c,p}}{c}\right)^2 t_p^2\, \mathbf{g} \mathbf{g}^T.
\end{equation}

\subsection*{A.4   Amplitude cross terms \texorpdfstring{$\mathbf{J}_{\tau\alpha}$}{J\_t,a} and \texorpdfstring{$\mathbf{J}_{\mathbf{v}\alpha}$}{J\_v,a} (Eq.\ (\ref{eq:Jtau_alpha_zero}))}
Using \eqref{eq:FIM_1} and \eqref{eq:FIM_3},
\begin{equation}
    \begin{aligned}
        \left(\frac{\partial \mu}{\partial \Re\{\alpha\}}\right)^{\!H}
\left(\frac{\partial \mu}{\partial \tau}\right)
=
\alpha\Big(\underbrace{\langle (d\odot a)\otimes s,\,(d\odot a)\otimes \partial_\tau s\rangle}_{=\langle s,\partial_\tau s\rangle=0}
+ \\ \underbrace{\langle (d\odot a)\otimes s,\,(d\odot \partial_\tau a)\otimes s\rangle}_{=\sum_p (-j2\pi f_{c,p})E_s}
\Big)
= -j2\pi \alpha\, F_1 E_s.
    \end{aligned}
\end{equation}
Likewise,
\begin{equation}
    \begin{aligned}
\left(\frac{\partial \mu}{\partial \Im\{\alpha\}}\right)^{\!H}
\left(\frac{\partial \mu}{\partial \tau}\right)
&= j\,\left(\frac{\partial \mu}{\partial \Re\{\alpha\}}\right)^{\!H}
\left(\frac{\partial \mu}{\partial \tau}\right) \\
&= \,2\pi \alpha\, F_1 E_s.
    \end{aligned}
\end{equation}
Taking real parts and arranging by the $(\Re\alpha,\Im\alpha)$ basis gives
\begin{equation}
    \begin{aligned}
\mathbf{J}_{\tau\alpha}
&=\frac{2}{\sigma_w^2}\Re\!\begin{bmatrix}
-j2\pi \alpha F_1E_s &   2\pi \alpha F_1E_s
\end{bmatrix} \\
&= \frac{4\pi E_s}{\sigma_w^2}F_1\,[\,-\Im\{\alpha^\ast\},\,-\Re\{\alpha^\ast\}\,].
    \end{aligned}
\end{equation}

Similarly, from \eqref{eq:FIM_1} and \eqref{eq:FIM_4},
\begin{equation}
    \begin{aligned}
\left(\frac{\partial \mu}{\partial \Re\{\alpha\}}\right)^{\!H}
\left(\frac{\partial \mu}{\partial v}\right)
&=\alpha\,\sum_p \langle (d\odot a)\otimes s,\,
(\partial_v d\odot a)\otimes s\rangle\\
&=\alpha\,\sum_p \langle d,\partial_v d\rangle\,\|s\|_2^2,
    \end{aligned}
\end{equation}
and $\langle d,\partial_v d\rangle=\sum_p \big(j2\pi t_p \tfrac{f_{c,p}}{c}\big)$, so
\begin{equation}
    \left(\frac{\partial \mu}{\partial \Re\{\alpha\}}\right)^{\!H}
\left(\frac{\partial \mu}{\partial v}\right)
= j2\pi \alpha\,\frac{E_s}{c}\,S_1\, \mathbf{g}^T,
\end{equation}
Likewise,
\begin{equation}
   \left(\frac{\partial \mu}{\partial \Im\{\alpha\}}\right)^{\!H}
\left(\frac{\partial \mu}{\partial v}\right)
= j\!\left(\frac{\partial \mu}{\partial \Re\{\alpha\}}\right)^{\!H}
\left(\frac{\partial \mu}{\partial v}\right)
= -2\pi \alpha\,\frac{E_s}{c}\,S_1\, \mathbf{g}^T. 
\end{equation}

Taking real parts yields
\begin{equation}
    \begin{aligned}
\mathbf{J}_{\mathbf{v}\alpha}
&=\frac{2}{\sigma_w^2}\Re\!\begin{bmatrix}
j2\pi \alpha \tfrac{E_s}{c}S_1 \mathbf{g}^T &   -2\pi \alpha \tfrac{E_s}{c}S_1 \mathbf{g}^T
\end{bmatrix} \\
&=\frac{4\pi E_s}{\sigma_w^2}S_1
\begin{bmatrix}
\Im\{\alpha\} & -\Re\{\alpha\}
\end{bmatrix} \mathbf{g}^T.
    \end{aligned}
\end{equation}

\subsection*{A.5   Delay–velocity cross block \texorpdfstring{$\mathbf{J}_{\tau \mathbf{v}}$}{J\_t,v} (Eq.\ (\ref{eq:delay_velocity_cross_term}))}
From \eqref{eq:FIM_3}\eqref{eq:FIM_4} the only nonvanishing contribution to
$\big(\partial_\tau \mu\big)^H\big(\partial_v \mu\big)$ comes from
$\big((d\odot\partial_\tau a)\otimes s\big)^H \big( (\partial_v d\odot a)\otimes s\big)$, since
$\Re\{\langle \partial_\tau s, s\rangle\}=0$ annihilates the term with $\partial_\tau s$:
\begin{align}
\left(\frac{\partial \mu}{\partial \tau}\right)^{\!H}
\left(\frac{\partial \mu}{\partial v}\right)
&= |\alpha|^2 \sum_p \big(\partial_\tau a_p\big)^\ast \big(\partial_v d_p\big) \, \|s\|_2^2\\
&= |\alpha|^2 E_s \sum_p \big(+j2\pi f_{c,p}\big)\,\Big(j2\pi t_p \frac{f_{c,p}}{c}\, \mathbf{g}^T\Big)\\
&= -\,|\alpha|^2 E_s \,\frac{4\pi^2}{c}\sum_p t_p f_{c,p}^2 \, \mathbf{g}^T .
\end{align}
Taking the real part and scaling by $2/\sigma_w^2$ as in \eqref{eq:FIM_complex} gives
\begin{equation}
\mathbf{J}_{\tau \mathbf{v}}
= -\,\frac{8\pi^2}{\sigma_w^2}\,|\alpha|^2 E_s \,\frac{1}{c}
\Big(\sum_p t_p f_{c,p}^2 \Big) \mathbf{g}^T.
\end{equation}

\section{Proof of Path Decoupling under Orthogonal Waveforms}
\label{appendix:orthogonality_proof}

We prove that if the transmitted waveforms are orthogonal in the delay–Doppler 
domain, then the FIM decouples across 
transmit–receive paths.

Consider the received baseband signal
\begin{equation}
\begin{aligned}
    y(t) &= \sum_{k=1}^{K} \mu_k(t; \theta_k) + w(t),
\\
\mu_k(t;\theta_k) &\triangleq \alpha_k\, s_k(t-\tau_k)\,e^{j2\pi \nu_k t},
\end{aligned}
\end{equation}
where $s_k(t)$ is the waveform of transmitter $k$, $\tau_k$ and $\nu_k$ are the 
propagation delay and Doppler shift of the corresponding path, 
$\alpha_k$ is an unknown complex amplitude, and $w(t)$ is zero-mean circular 
complex Gaussian noise.

Define the cross-ambiguity function between two waveforms $s_k$ and $s_m$ as
\begin{equation}
A_{km}(\tau,\nu) 
\triangleq \int_{-\infty}^\infty s_k(t)\, s_m^*(t-\tau)\,e^{-j2\pi \nu t}\,dt.
\end{equation}
We define the waveforms to be orthogonal over a domain $\mathcal D$ if
\begin{equation}
A_{km}(\tau,\nu) = 0, 
\quad \forall (\tau,\nu)\in\mathcal D,  \forall k\neq m.
\label{eq:DD_orth}
\end{equation}

The log-likelihood for parameters $\theta = \{\tau_k,\nu_k,\alpha_k\}$ has 
FIM entries
\begin{equation}
J_{ij} = \frac{2}{N_0}\,\Re\!\left\{
\left(\frac{\partial \mu}{\partial \theta_i}\right)^H
\left(\frac{\partial \mu}{\partial \theta_j}\right)
\right\}.
\label{eq:FIM_general}
\end{equation}

Now consider the cross-path term with $\theta_i$ from path $k$ and $\theta_j$ from path $m\neq k$.
Then the cross-path information is
\begin{equation}
J_{ij} = \frac{2}{N_0}\,\Re\!\left\{
\left(\frac{\partial \mu_k}{\partial \theta_i}\right)^{\!H}
\left(\frac{\partial \mu_m}{\partial \theta_j}\right)
\right\}.
\label{eq:cross_path}
\end{equation}

For each path $k$,
\begin{align}
\frac{\partial \mu_k}{\partial \tau_k} &= -\alpha_k \,\dot s_k(t-\tau_k)\,e^{j2\pi \nu_k t},\\
\frac{\partial \mu_k}{\partial \nu_k} &= j2\pi t\,\alpha_k \, s_k(t-\tau_k)\,e^{j2\pi \nu_k t},\\
\frac{\partial \mu_k}{\partial \alpha_k} &= s_k(t-\tau_k)\,e^{j2\pi \nu_k t},
\end{align}
where $\dot s_k(\cdot)$ denotes the time derivative of $s_k(\cdot)$. Each derivative is thus a linear combination of time- and frequency-shifted 
versions of $s_k$, or its time derivative.

Substituting these derivatives into \eqref{eq:cross_path}, all cross-path inner 
products reduce to linear combinations of cross-ambiguity terms 
$A_{km}(\tau_k-\tau_m,\nu_k-\nu_m)$ or their $\tau$-derivatives.  
By assumption \eqref{eq:DD_orth}, and provided the uncertainty domain is 
contained in $\mathcal D$, these cross-ambiguities vanish identically.  
Therefore,
\begin{equation}
J_{ij} = 0 \qquad \forall \text{$i,j$ s.t. $\theta_i\in\theta_k$, $\theta_j\in\theta_m$, $k\neq m$.}
\end{equation}
The FIM has no cross-path blocks and is thus block diagonal across paths:
\begin{equation}
\mathbf{J}(\boldsymbol{\theta})
 = 
\operatorname{diag}\big(\mathbf{J}_1,\mathbf{J}_2,\ldots,\mathbf{J}_k\big),
\label{eq:FIM_blockdiag}
\end{equation}
where each $\mathbf{J}_k$ is the Fisher information corresponding to path $k$. 
This establishes rigorously that orthogonal waveforms decouple the estimation 
problems of distinct paths.
\end{appendices}

\bibliographystyle{IEEEtran}
\bibliography{main}

\end{document}